\documentclass[aps,onecolumn]{revtex4}
\usepackage{amsmath}
\usepackage{makeidx}
\usepackage{graphicx}
\usepackage{amsfonts}
\usepackage{amssymb}

\begin{document}

\newtheorem{defn}{Definition}
\newtheorem{theo}{Theorem}
\newtheorem{ex}{Example}
\newtheorem{prop}[theo]{Proposition}
\newtheorem{lemma}[theo]{Lemma}
\newtheorem{coroll}[theo]{Corollary}
\newtheorem{conj}[theo]{Conjecture}
\newtheorem{note}[defn]{Point}
\newtheorem{apprx}[defn]{\underline{Approximation}}
\newcommand{\fine}{ $ \Box  $}

\title{Statistical Mechanics of maximal independent sets}
\author{Luca Dall'Asta}
\affiliation{
The Abdus Salam International Centre for Theoretical
Physics, Strada Costiera 11, 34014 Trieste, Italy}
\author{Paolo Pin}
\affiliation{%
Dipartimento di Economia Politica, Universit\'a degli Studi di Siena, Piazza San Francesco 7, 53100 Siena, Italy}
\author{Abolfazl Ramezanpour}
\affiliation{Dipartimento di Fisica, Politecnico di Torino, Corso Duca degli Abruzzi 24, 10129 Torino, Italy}

\begin{abstract}
The graph theoretic concept of maximal independent set arises in several practical problems in computer science as well as in game theory.
A maximal independent set is defined by the set of occupied nodes that satisfy some packing and covering constraints. It is known that finding
minimum- and  maximum-density maximal independent sets are hard optimization problems. In this paper we use cavity method of statistical physics
and Monte Carlo simulations to study the corresponding constraint satisfaction problem on random graphs. We obtain the entropy of maximal
independent sets within the replica symmetric and one-step replica symmetry breaking frameworks, shedding light on the metric structure of the landscape of solutions and suggesting a class of possible algorithms.  This is of particular relevance for the application to the study  of strategic interactions in social and economic networks, where maximal independent sets correspond to pure Nash equilibria of a graphical game of public goods allocation.
\end{abstract}

\maketitle

\section{Introduction}
It is well known that an important family of computationally difficult problems are concerned with graph theoretical concepts, such as covering and packing \cite{K72}. Among them, the problem of finding a {\em minimum vertex cover} has probably become the typical example of  NP-hard optimization problems defined on graphs \cite{GJ79} and it has recently attracted a lot of attention in the statistical physics community for its relation with the physics of spin glasses \cite{HWbook}.
In particular, in statistical mechanics the vertex cover problem is usually studied in its dual representation of hard-core lattice gas, where coverings are mapped into particles of unit radius that cannot be located on neighboring vertices.
In graph theory, such a dual configuration defines an {\em independent set}, i.e. a set of vertices in a graph no two of which are adjacent.
More formally, given a graph $\mathcal{G} = (\mathcal{V},\mathcal{E})$ a subset $\mathcal{I} \subseteq\mathcal{V}$ of vertices is an {\em independent set} if for every pair of vertices $i,j \in \mathcal{I}$ the edge $(i,j) \not\in \mathcal{E}$.
The {\em size} of an independent set is the number of vertices it contains. So, given a graph $\mathcal{G}$ with $N$ vertices and an integer $M < N$ it is reasonable to ask if it is possible to find an independent set of size at least $M$. This decisional problem was proved to be NP-complete and its optimization version, i.e. finding an independent set of the largest size ({\em maximum independent set}), is NP-hard \cite{K72,GJ79}.
From the definition it is straightforward to notice that the complement of an independent set is a vertex cover and finding a minimum vertex cover is exactly equivalent to find a maximum independent set.

\begin{figure}
\includegraphics[width=8cm,angle=0]{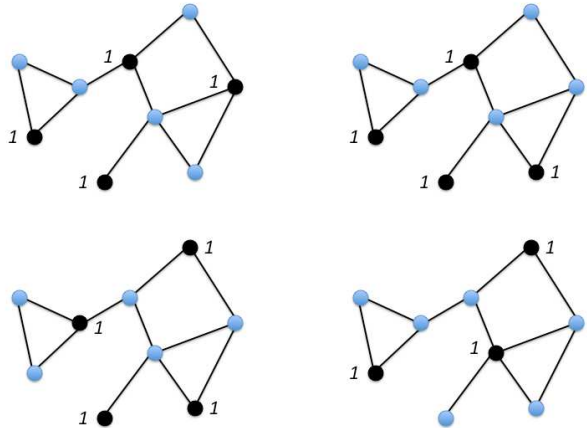}
\caption{(Color online) Example of 4 maximal-independent sets (mIS) out of the 11 possible ones for this graph with 9 vertices. Vertices labeled with $1$ belong to the mIS.
} \label{diag1}
\end{figure}

We are here interested to a slightly different graph theoretic problem, dealing with the concept of {\em maximal independent set} (mIS),  that is an independent set that is not a subset of any other independent set.
This means that adding a node to a maximal independent set would force the set to contain an edge, contradicting the independence constraint. Again, by removing a vertex from an independent set we still get an independent set but the same is not true for maximal independent sets. In this sense, mISs present some property typical of packing-like problems that makes the corresponding optimization problem quite different from the widely studied vertex cover problem.
In particular, it turns out that mISm are actually the intersection of independent sets and vertex covers.
 See Fig. \ref{diag1} for some examples of mISs in a small graph.

A maximal independent set can be easily found in any graph using simple greedy algorithms, but they do not allow to control the size of the mIS. As for the independent set problem, the complexity increases if we are asked to find maximal independent sets of a given size $M$ and, in particular, the problems of finding maximal independent sets of maximum and minimum size are NP-hard. Note that as well as a maximum maximal independent set (MIS) (maximum independent set) is equivalent to a minimum vertex cover, a minimum maximal independent set (mis) is often addressed as {\em minimum independent dominating set} \cite{GJ79}.

Apart from the purely theoretical interest for problems that are known to be computationally difficult, finding maximal independent sets plays an important role in designing algorithms for studying many other computational problems on graphs, such as $k$--Coloring, Maximal Clique and Maximal Matching problems.
Moreover, distributed algorithms for finding mISs, like Luby's algorithm \cite{L86}, can be applied to networking, e.g. to define a set of mutually compatible operations that can be executed simultaneously in a computer network or to set up message-passing based communication systems in radio networks \cite{MW05}.
One recent and interesting application is in microeconomics, where maximal independent sets can be identified with {\em Nash equilibria} of
a class of network games representing public goods provision \cite{BK07,GGJVY08}. Hence, studying maximal independent sets allows to understand the properties of Nash equilibria in these network games.

Methods from statistical mechanics of disordered systems turned out to be extremely effective in the study of combinatorial optimization and computational problems, in particular in characterizing the ``average case'' complexity of these problems, i.e. studying the typical behavior
of randomly drawn instances (ensembles of random graphs) \cite{MPZ02} that can be very different from the worst case usually analyzed in theoretical computer science.
Following the standard lattice gas approach, we represent maximal independent sets as solutions of a constrained satisfaction problem (CSP) and provide a deep and extensive study of their organization in the space $\{ 0, 1\}^N$ of all lattice gas configurations.

On general graphs the number of maximal independent sets is exponentially large with the number of vertices $N$, therefore an interesting problem is to compute their number as a function of their size $M$.
On random graphs, this number can be evaluated using different methods. An upper bound is obtained analytically using simple
combinatorial methods, whereas a more accurate estimate is given by means of the replica symmetric cavity method.
These methods allow to compute the entropy of solutions (i.e. of maximal independent sets) as a function of the density of coverings $\rho = M/N$ and are approximately correct in the large $N$ limit.

There are density regions in which no maximal independent set can be found, that correspond to the UNSAT regions of the phase diagram of the associated CSP. We study the structure of this phase diagram as a function of the average degree of the graph ($K$ for random regular graphs and $z$ for Erd\"os-R\'enyi random graphs).
Since the replica symmetric (RS) equations (belief propagation) are not always exact, we study their stability in the full range of density values $\rho$ and discuss  the onset of replica symmetry breaking (RSB). This is expected, because both problems of finding minimum and maximum mISs are NP-hard. The two extreme density regions are not symmetric and different behaviors are observed; in particular
while  RS  solutions are stable in the low density region they are unstable in the large density region.  For random regular graphs with connectivity $K=3$,  general one-step replica symmetry breaking (1RSB) studies show a  dynamical transition in the low density region and a condensation transition at high densities (accompanied by a dynamical one). In the latter case, the complexity remains zero in both the RS and 1RSB phases.

The theoretical results valid on ensembles of graphs, are tested on single realizations using different types of algorithms.
Greedy algorithms converge very quickly to maximal independent sets of typical density, around the maximum of the entropic curve.
By means of Monte Carlo methods, as well as message passing algorithms, we can explore a large part of the full range of possible density
values. As expected, finding maximal-independent sets becomes hard in the low and high density regions, but each algorithm
stops finding mISs at different density values.

The paper is organized as follows: In the next section we recall some known mathematical results and discuss related works in computer science, economics, and physics;
Section III is instead devoted to present some rigorous results on the spatial organization of maximal independent sets (in their lattice gas representation).
We develop the cavity approach in Sec. IV, with both the RS solution and the discussion of the replica symmetry breaking in regular random graphs.
From Sec. V we move our attention to the problem of  developing algorithms to find maximal-independent sets of typical and non-typical size (i.e. density $\rho$). We present a detailed analysis of greedy algorithms, and put forward several different Monte Carlo algorithms, whose  performances are compared with one of the best benchmarks used in CSP analysis, the Belief Propagation-guided decimation.
Conclusions and possible developments of the present analysis with application to computer science and game theory are presented in Section VI.\\
Some more technical results are reported in the Appendices. In Appendix A we compute upper bounds for the entropy of maximal independent sets in random regular graphs and Poissonian random graphs using a combinatorial approach in the annealed approximation (first and second moment methods); while Appendix B contains a proof of the results in Section \ref{orgSEC}.
In Appendix C we discuss the Survey Propagation and criticize its results. Appendix D is devoted to the important calculations for the RS and 1RSB stability analysis. In Appendix E we shortly describe the Population Dynamics algorithm that is used
to solve some equations in the paper. In Appendix F we analyze separately the two constraints defining a mIS, i.e.  {\em neighborhood covering} and  {\em hard-core particle packing} problems that approximate (from above) the correct entropy in the low  and  high density regions respectively.
Finally, in Appendix G we briefly consider the problem of enumerating mISs of higher order, that is also relevant to study stable specialized Nash equilibria of some network games \cite{BK07}.

\section{Related Work and known results}
This section is devoted to collect and briefly summarize a large quantity of results and works directly or indirectly
related to maximal independent sets in graphs. For simplicity we have divided them depending on their fields of application.

\subsection{Computer Science and Graph Theory}

A first upper bound for the total number of  maximal independent sets was obtained by Moon and Moser (1965) showing that any graph with $N$ vertices has at most $3^{N/3}$ mISs \cite{MM65}. The disjoint union of $N/3$ triangle graphs is the example of a graph with exactly $3^{N/3}$.
On a triangle-free graph, the number of mISs is instead bounded by $2^{N/2}$ \cite{HT93}.

In case we are interested to the number of mISs of a fixed size $M$, tighter bounds are available.
In relation to coloring, it was shown by Byskov \cite{B03} that the number of mISs of size $M$ in any graph of size $N$ is at most
\begin{equation}
{\left \lfloor \frac{N}{M}\right \rfloor}^{M-(N\mod{M})} {\left \lfloor \frac{N}{M}+1 \right \rfloor}^{N\mod{M}}
\end{equation}
while the number of mISs of size $\leq M$ is at most $3^{4M-N} 4^{N-3M}$\cite{E03}. These upper bounds can be compared  with those presented here and obtained using non-rigorous methods (see Tables \ref{table1}-\ref{table2}).

Despite these mathematical results, counting maximal independent sets is mainly a computational problem. In particular, listing all mISs of a given graph and retaining the ones of maximum and minimum size allows to solve the maximum independent set and minimum independent dominating set problems, that are NP-hard. Again, a list of all mISs can be exploited to find $3$-coloring of a graph and to solve other difficult problems \cite{L76}. For this reason, researchers have studied algorithms to list all maximal independent sets efficiently in polynomial time per output set, i.e. polynomial delay between two consecutively produced mISs \cite{misall}. Even faster is the celebrated randomized distributed algorithm proposed by Luby, and based on message-passing technique, that works in $\mathcal{O}(\log N)$ time \cite{L86}.
The main problem is that the number of mISs is in general exponential with the number of vertices $N$, therefore the total time required to list all maximal independent sets is still too large for computational purposes.

Another important result is the {\em inapproximability} within any constant or polynomial factor of the associated optimization problem. Indeed, a branch of computer science is interested in proving the existence of algorithms that find in polynomial time suboptimal solutions to optimization problems that are NP-hard. These algorithms can be proved to be optimal up to a small constant or a polynomial factor (see e.g. \cite{V03}).
When this can be done, the optimization problem is approximable. For the minimum  vertex cover it is possible to find a cover that is at most
twice as large as the optimal one, therefore MVC is approximable with constant factor $2$. On the contrary  the maximum mIS problem (i.e. finding a MVC configuration but maintaining the ``maximality'' condition at every step) is not approximable within any constant or polynomial factor \cite{FGLSS91}. This result shows that maximality condition is not a minor detail and actually has a profound impact on the properties of the computational problem.

\subsection{Economics and Game theory}

The lattice gas configurations associated with maximal independent sets are Nash Equilibria of a discrete network game, called Best Shot Game, recently studied by Galeotti et al. \cite{GGJVY08} and previously introduced for the case of complete networks by Hirshleifer \cite{H83}.
The Best-Shot game is a toy-model for the type of strategic behaviors that emerges in many social and economic scenarios ranging from information collection to public-goods, i.e. wherever agents are allowed to free ride and exploit the actions of their peers.

A more general game theoretic framework for the allocation of public goods on a network structure was proposed by Bramoull\'e and Kranton \cite{BK07}.
In their game, agents are located on the nodes of a network and have to decide about the investment of resources for the allocation of some public goods. An agent can purchase the good for a fixed cost $c$, cooperate with neighboring agents sharing a lower individual investment (i.e. $< c$),  or free ride possibly exploiting a neighbor's investment.  Two classes of equilibria exist: {\em specialized equilibria}, in which agents either pay all cost $c$ or free ride, and {\em mixed equilibria} where cooperation is present. However only specialized equilibria are stable and for this reason we can  limit the analysis to the discrete case with only two actions: action 1 (full cost investment) and action 0 (free-riding) \cite{GGJVY08}.
In the Best Shot game agents possible actions are restricted to these two options.
Independent sets arise because agents receive positive externalities, i.e.  they prefer not to contribute if at least one of their neighbors already does.
The independent sets are maximal because the contribution is a \emph{dominant action}, i.e. one is better off by contributing if none of the neighbors does.
The set of Nash equilibria of the Best Shot game on a graph $\mathcal{G}$ is exactly the set of maximal-independent sets of $\mathcal{G}$.

In the case of public goods the objective function of a Social Planner is to find optimal Nash equilibria, that are Nash equilibria maximizing the sum of individual utilities by minimizing the global cost.
We have seen that finding MIS and mis is an NP-hard optimization problem, therefore we expect that self-organizing towards such optimal
equilibria is an equally difficult task for a network of economic agents. It is thus of great importance to develop simple mechanisms to trigger optimization and study  how such mechanisms could be implemented in realistic situations for instance by means of economic incentives \cite{DPR09}.

It is worth noting that recently some techniques from statistical physics have been applied to the study of the Best Shot game \cite{LP07}. The work by Lopez-Pintado \cite{LP07}, indeed, put forward a mean-field theoretical analysis of the best-response dynamics that provides an estimate of the average density of contributors (action $1$) in Nash Equilibria on general uncorrelated random graphs. In Section \ref{numSEC}, we will discuss the relation between best-response dynamics and another algorithm (the Gazmuri's algorithm) that can be used to find Nash Equilibria (i.e. mIS).

\subsection{Physics}
The hardcore lattice gas representation allows to map maximal independent sets on the solutions of a constraint satisfaction problem.
As mentioned in the introduction, similar CSPs have been recently studied in the statistical mechanics community to model
systems with geometrical or kinetic constraints, and exhibiting a glass transition.

Kinetically constrained models \cite{RS03} are used as simple, often solvable, examples of the glass transitions. For instance, the Kob-Andersen model \cite{KA93} is a lattice gas with a fixed number of particles in which a particle is mobile if the number of occupied neighbors is lower than a given threshold, mimicking the {\em cage effect} observed in super-cooled liquids. At sufficiently large density of particles, the system can be trapped in some blocked configuration, in which all particles are forbidden to move.
Though the model is dynamical and satisfies constraints of a kinetic nature, the number of blocked states depending on the parameters of the system can be studied with a purely static, thermodynamic approach ({\em \'a la Edwards}) \cite{E94}.
These calculations, performed using the transfer matrix methods, the replica method and numerically by thermodynamic integration, have been applied to several models exhibiting dynamical arrest, such as the Fredrickson-Andersen model \cite{FA84}, the zero-temperature Kawasaki exchange dynamics \cite{DGL02,DGL03} or urn models \cite{R95}, and recently extended to the study of Nash Equilibria of the Schelling's model of segregation \cite{DCM08}.

Studying the dynamical arrest only by means of thermodynamic methods leads to underestimate the role of the dynamical basins of attraction of different blocked configurations (Edwards measure vs. dynamical measure).
On the contrary, these methods become exact or approximately correct in problems with constraints due to geometric frustration, like
hard-core lattice gas models \cite{WH03} and hard-sphere packing problems \cite{PZ08}.
As mentioned in the Introduction, the simplest hard-core lattice gas model is the dual of the vertex covering where the close-packing limit (high density regime) corresponds exactly to the minimum vertex cover problem (i.e. maximum independent set problem).
On random graphs with average degree $z > e$, the vertex covering problem presents frustration and long-range correlations, that are  responsible of replica symmetry breaking \cite{Z05}. Moreover,
the numerical detection of hierarchical clustering in the solution's landscape suggests the existence of levels of RSB higher than 1RSB \cite{BH04}.

A first attempt to model a purely thermodynamically driven glass transition in a particle system was proposed by Biroli and M\'ezard (BM), that studied a {\em lattice glass model} on regular lattices \cite{BM02} and random regular graphs \cite{RBMM04}, in which configurations violating some locally defined density constraints are forbidden. More precisely, a particle cannot have more than $\ell$ among its $k$ neighbors occupied.
At zero temperature, the constraints become hard and the model is effectively a CSP defined on a general graph.
The BM model is  very similar to our problem as shown by mapping back particles onto uncovered vertices and vacancies
onto covered ones. In a configuration defining a maximal independent set, every uncovered node has at most $k-1$ uncovered neighbors, because at least one of them has to be covered. Therefore, our model is similar to a BM model with $\ell = k-1$.
However, a further constraint on covered nodes is present, requiring vacancies to be completely surrounded by particles.
Maximal independent sets are thus similar to lattice glass configurations in the close packing regime, but the existence of the packing-like  constraint prevents the statistical mechanics of the two models to be exactly the same. A generalization of BM model with attractive short rang interactions has been studied in Ref. \cite{KTZ08}.

Finally, a recent paper by Tarzia and M\'ezard \cite{TM07} puts forward a model of {\em Hyper Vertex Covering} (HVC) that is an abstraction of the Group Testing procedures. The model can be defined on a random regular bipartite factor graph with $N$ variables and $M$ constraints and consists in finding a cover of the variables (that are $1$ or $0$ if covered or uncovered respectively) subject to the condition that the sum of the variables involved in each constraint is at least $1$. This hypervertex cover constraint is somewhat similar to the maximal-independent set constraint that requires at least one of the nodes in the set composed by a node and its neighborhood to be $1$. On the other hand,  maximal-independent sets are defined on real graphs instead of factor graphs and there is a further packing-like constraint (no neighboring 1s are allowed).

In conclusion, similar problems are current matter of investigation in the statistical physics community \cite{BBMTWZ}, but in our opinion the mIS  problem is
different from all them and substantially new due to the presence of two local constraints with contrasting effects.

\section{Maximal Independent Sets: existence and organization}
\label{orgSEC}

A preliminary account of the statistical properties of maximal independent sets can be obtained from simple but rigorous mathematical analyses providing {\em i)} {\em lower and upper bounds} for the existence of mISs with density $\rho$ of covered nodes, and {\em ii)} information on their {\em spatial arrangement} in the space of all possible binary configurations on a graph.

We have anticipated in the previous sections that in general it is not possible to find maximal independent sets
at any given density $\rho$ of covered nodes. This is due to the conflicting presence of covering-like constraints (favoring higher densities) and packing-like constraints (favoring lower densities).  We thus expect to find mISs only within a finite range of density values $[\rho_{min}, \rho_{max}]$. On random graphs, lower bounds $\rho_{min}^{lower}$ for the minimum density value and upper bounds $\rho_{max}^{upper}$
for the maximum density value can be easily obtained by means of annealed calculations employing the first moment method (see Appendix \ref{appANN}). These values are reported in Table \ref{table1} for the case of Erd\"os-R\'enyi random graphs and in Table \ref{table2} for random regular graphs (RRG), i.e. graphs in which connections are established in a completely random way with the only constraint that all nodes have the same finite degree $K$.

In addition to the results of these combinatorial methods, very general results on the structure and organization of mISs
can be obtained rigorously starting from the lattice gas representation. (We refer to the Appendix \ref{appORG} for the proofs of all propositions contained in the present section.)
An independent set $I$ is a configuration with binary variables $\sigma_i \in \{0,1\}$, in which $\sigma_i = 1$ if the node $i$  belongs to the independent set ($i \in I$) and $\sigma_i = 0$ if the node is not part of it ($i \in I^c$). We are interested only in those independent sets that are maximal and we focus on the behavior of single variables as well as sets of variables. We first introduce some useful concepts.

Consider a set $\mathcal{S}$ of solutions of the mIS problem, that is a set of configurations $\underline{\sigma}$, with the property of being a mIS for a given graph $\mathcal{G}$.
A variable $\sigma_i$ is {\em frozen} in the set $\mathcal{S}$ if it is assigned the same value in all  configurations belonging to $\mathcal{S}$.
Therefore, by flipping this variable we cannot obtain a configuration with the property of being in the same set $\mathcal{S}$.
For matter of convenience we classify these kind of variables in a hierarchical way. The variable $\sigma_i$ is {\em locally frozen} if we can obtain a configuration belonging to $\mathcal{S}$ by flipping $\sigma_i$ and at most a number $n$ of other variables $\sigma_j$ with strictly $n = o(N)$.
The variable $\sigma_i$ is instead {\em globally frozen} if, in order to have another configuration in $\mathcal{S}$, we have to flip $\sigma_i$ and $\mathcal{O}(N)$  other variables.

In the following, we consider $\mathcal{S}$ to be the set of all maximal independent sets in a given graph $\mathcal{G}$.
It is straightforward to show that (see App. \ref{appORG})

{\prop \label{prop1} If a configuration is a maximal independent set, then all variables $\sigma_{i}$ $\forall i =1,\dots N$ are at least locally frozen.}\\

Proposition \ref{prop1} shows that the Hamming distance between two mIS configurations is at least $2$, but does not specify if this is the actual minimum in every graph and, more importantly, which is the maximum possible distance between two mISs.
 A result in this direction is obtained by considering the propagation of variable rearrangements induced by a single variable flip. Suppose that at step $t=0$ a variable $\sigma_i$ is flipped from $0$ to $1$, then some of the local constraints on the neighboring variables become unsatisfied and these variables have to be flipped too. At step $t=1$, we flip all variables giving contradictions and identify all other constraints that now become unsatisfied. We proceed in this way until all variables satisfy their local constraints and the configuration is again in $\mathcal{S}$. This iterative operation is usually called {\em best-response} dynamics in game theory \cite{GGJVY08}. In many CSPs,  a similar iteration would not converge rapidly to a new solution due to the presence of loop-induced frustration and long-range correlations. In such cases, variables are globally frozen, because their value depends on the value assumed by an extensive number of other variables. On the contrary, in the case of maximal independent sets, all rearrangements involve a finite number of variables.

{\prop \label{prop3} If a  variable $\sigma_i$ of a mIS is flipped and the variables (including $\sigma_i$ itself) are updated just once by best-response dynamics, the outcoming configuration is still a maximal independent set. Moreover:
A) If a variable $\sigma_i$ of a mIS is flipped from $1$ to $0$ the best-response dynamics is limited to the neighborhood of the node $i$.
B) If instead the variable $\sigma_i$ is flipped from $0$ to $1$, the best-response dynamics extends at most to the second neighborhood of $i$.}\\

\begin{figure}
\includegraphics[width=10cm,angle=0]{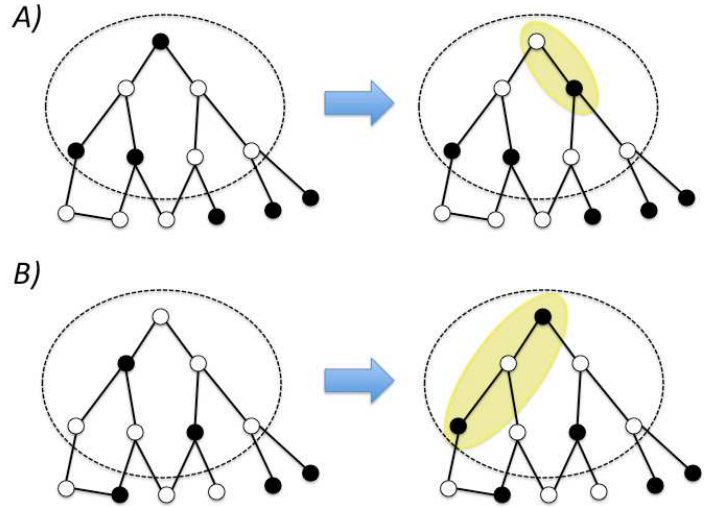}
\caption{(Color online) Examples of rearrangements induced in a mIS configuration by a single spin flip: A) from $1$ (black) to $0$ (white), B) from $0$ to $1$.} \label{diag2}
\end{figure}

An example of the validity of Prop. \ref{prop3} is reported in Fig. \ref{diag2}. In A) we flip the top vertex from $1$ (black) to $0$ (white) and rearrange the rest of the mIS configuration. The right neighbor is forced to flip $0 \to 1$, but her neighbors are $0$ thus they do not flip. The rearrangement propagates up to the neighbors of the top node.  In B) the top node is flipped from $0$ to $1$. The left neighbor flips to $0$, thus leaving the neighboring node unsatisfied. This node flips to $1$, but all her neighbors are already satisfied at $0$. The rearrangement propagates here up to the second neighborhood.

The number of  variables flipped during the best-response dynamics depends on the topological properties of the underlying graph.
When the first and the second moments of the degree distribution are finite, Prop. \ref{prop3} implies that all variables are only locally frozen, and starting from a mIS it is always possible to find another mIS within a finite number of spin flips.
This is not always true for scale-free networks with power-law degree distribution $p_k \propto k^{-\gamma}$ and $\gamma <3$. In these networks, the second moment of $p_k$ diverges with the system's size $N$, therefore a single variable flip can induce the rearrangement by best-response of an extensive number of other variables. An example is provided by star-like graphs in which the only two possible maximal independent sets are the one with the central node covered and all other nodes uncovered and the complementary configuration. It is easy to see that the best-response dynamics following a single flip always extends to the whole graph and implies the rearrangements of $\mathcal{O}(N)$ variables, that are globally frozen.

Let us exclude extremely heterogeneous star-like structures, by focusing on homogeneous graphs with bounded degrees, i.e. $\forall i$ $k_i < const \ll N$.
Even in  case of networks in which maximal independent sets are locally frozen, these mIS are all connected by best response dynamics, as proven by the following result.

{\prop \label{prop4} Given a pair of maximal independent sets $\mathcal{I}$ and $\mathcal{I}'$, it is always possible to go from $\mathcal{I}$ to $\mathcal{I}'$ with a finite sequence of operations, each one consisting in flipping a single variable from $0$ to $1$ followed by a best-response rearrangement.} \\

This statement proves exactly the {\em connectedness} of the space of maximal independent sets under an operation that allows to move on that space. In the thermodynamic limit, when the distance between two mISs can be taken of $\mathcal{O}(N)$, there is a chain of at most $\mathcal{O}(N)$ operations to go from one mIS to the other, and each intermediate step corresponds to a mIS that differs from the previous and the following ones in just a finite number of variables (due to a finite rearrangement).

In summary, we have uncovered two important properties of the space of mISs: 1) all variables in a mIS are locally and not globally frozen, i.e.
the minimum Hamming distance between mISs is at least $2$;
2) the space of mIS is connected if we move between mIS using a well-defined sequence of operations.
It means that all mISs form a single ``coarse-grained'' cluster when the appropriate scale is chosen, that is that of best response rearrangement operation.  \\
{\em How does this picture change when we take into account mISs at fixed density $\rho$ of occupied nodes?} It is not clear, because
in the rearrangement that follows each single variable's flip, the number of covered nodes changes of a finite amount, causing only a negligible variation in the density $\rho$. On the other hand, the length of the sequence of these operations can be of $\mathcal{O}(N)$, therefore two mISs at the same, very low or very high, density could be connected by a very long path that gradually overpasses some density barrier.
Since the proof of  connectedness cannot be easily extended to the space of mISs at fixed density $\rho$, we cannot exclude the onset of some clustering phenomena at very low or very high values of $\rho$.
In the next section we will try to address this question by means of statistical mechanics methods.

\begin{table}
\begin{center}

\begin{tabular}{|c||c|c||c|c|}
  \hline

$z$      &  $\rho_{min}^{lower}$ &  $\rho_{min}^{BP}$  & $\rho_{max}^{BP}$ & $\rho_{max}^{upper}$ \\
  \hline

$3$      & $0.253$ & $0.301$ & $0.561$  & $0.631$    \\

  \hline

$4$      & $0.216$ &  $0.244$ & $0.504$  & $0.564$  \\

  \hline

$5$      & $0.190$ &  $0.208$ &  $0.461$   & $0.511$  \\

  \hline

$6$      & $0.170$ &  $0.183$ &  $0.425$   & $0.468$  \\

  \hline

$7$      & $0.155$ &  $0.165$ & $0.396$   & $0.432$  \\

  \hline

$8$      & $0.143$ &  $0.150$ &  $0.370$   & $0.403$  \\

  \hline

$9$      & $0.132$ &  $0.138$ & $0.350$   & $0.377$  \\

  \hline

$10$      & $0.124$ & $0.129$ &  $0.331$  & $0.355$ \\

  \hline

\end{tabular}

\vskip 0.5cm

\caption{Erd\"os-R\'enyi Graphs of average degree $z$: Comparing lower and upper bounds $\rho_{min}^{lower}$,
  $\rho_{max}^{upper}$, obtained with the first moment method, with
  the values that Belief Propagation (BP, see Section V) equations predict $\rho_{min}^{BP}$,
  $\rho_{max}^{BP}$.}\label{table1}
\end{center}
\end{table}

\section{Phase diagram by the cavity method}\label{SECcavity}

Some of the information obtained in the previous section are now
compared with the results of a statistical mechanics analysis \cite{MPV87,MM09} by means of
the zero-temperature cavity method \cite{MP02}.
At the level of graph ensembles, there is a sharp difference between
random regular graphs and ER random graphs.
In random regular graphs, all nodes behave in the same way, thus the
cavity equations simplify considerably leading to simple recursion
equations for a small set of probability marginals.
The case of ER random graphs is more involved as the presence of
various degree values implies the use
of distributions instead of simple cavity fields already at the
replica-symmetric (RS) level. In both classes the
solutions of the replica symmetric cavity equations are not always stable
(at least for some values of the average degree).
For random regular graphs we consider the 1-step replica symmetry
breaking (1RSB) scenario to see how glassy phases (if any) change the RS picture
close to minimum and maximum densities.

However, cavity equations can be applied to study single instances as
well \cite{MeZ02}, leading to message-passing recursive algorithms
able to find efficiently maximal independent sets in a wide range of
density values on very general kinds of graphs (see Section \ref{BPdecSEC}).

\begin{table}
\begin{center}

\begin{tabular}{|c||c|c|c||c|c|c|}
  \hline

    $K$     & $\rho_{min}^{lower}$ &  $\rho_{min}^{BP}$ &
    $\rho_{s_1}^{BP}$ &  $\rho_{s_2}^{BP}$ & $\rho_{max}^{BP}$   & $\rho_{max}^{upper}$  \\
  \hline

    $3$     & $0.233$  & $0.264$ & $-$ & $0.425$ & $0.458$  & $0.649$  \\

  \hline

    $4$    &  $0.20$  & $0.223$ & $-$ & $0.381$ & $0.419$  & $0.578$  \\

  \hline

    $5$    &  $0.177$ & $0.196$ & $-$ & $0.349$ & $0.387$  & $0.522$ \\

  \hline

    $6$    & $0.160$  & $0.175$ & $-$  & $0.324$  & $0.360$ & $0.476$  \\

  \hline

    $7$     & $0.147$ & $0.159$ & $0.166$   & $0.303$   & $0.338$ & $0.439$ \\

  \hline

    $8$      & $0.135$ & $0.146$ & $0.157$  & $0.285$  & $0.319$  & $0.408$ \\

  \hline

    $9$     & $0.126$  & $0.136$ & $0.150$  & $0.272$  & $0.301$ & $0.382$ \\

  \hline

    $10$    & $0.118$  & $0.127$ &  $0.144$  &  $0.258$  & $0.287$  & $0.359$ \\

  \hline

\end{tabular}

\vskip 0.5cm

\caption{Random Regular Graphs of degree $K$: Comparing lower and upper bounds $\rho_{min}^{lower}$,
  $\rho_{max}^{upper}$, obtained with the first moment method, with
  the values that Belief Propagation (BP, see Section V) equations predict $\rho_{min}^{BP}$,
  $\rho_{max}^{BP}$. For $\rho<\rho_{s_1}^{BP}$ we need damping to converge BP equations while for
$\rho>\rho_{s_2}^{BP}$ BP equations do not converge even with damping.}\label{table2}
\end{center}
\end{table}

\subsection{Replica Symmetric results: belief propagation}
The problem of finding
a maximal independent set of density $\rho$ can be mapped on the
 problem of finding the ground state of a
particular kind of spin model or lattice gas on the same graph.
On each vertex $i$ we define a binary variable $\sigma_i =
\{0,1\}$. For a configuration $\underline{\sigma} = \{\sigma_i |
i = 1, \dots, N \}$ to be in the ground state (i.e. a mIS), each
variable $i$ has to satisfy a set of $k_i +1$ constraints involving
neighboring variables.
There are $k_i$ constraints $I_{ij}$ on the edges emerging from $i$, each one involving two neighboring variables $i$ and $j$, and one further constraint $I_i$ on the whole neighborhood of $i$. For two neighboring nodes $i$ and $j$, the edge constraint $I_{ij} = 1$ iff $\sigma_i  = 0 \vee \sigma_j = 0$ ({\em packing-like constraint}); while the neighborhood constraint $I_{i} = 1$ iff $\sigma_i + \sum_{j \in \partial i} \sigma_j >0$ ({\em covering-like constraint}). Here $\partial i$ represents the set of neighbors of node $i$.

The zero temperature partition function corresponding to this constraint-satisfaction problem reads
\begin{equation}
Z(\mu) = \sum_{\underline{\sigma}} \prod_{i}  I_i(\sigma_{i},\sigma_{\partial i}) \prod_{(i,j)\in \mathcal{E}} I_{ij}(\sigma_i,\sigma_j) e^{-\mu \sum_i \sigma_i}
\end{equation}
in which $\sigma_{\partial i} = \{\sigma_j | j \in \partial i \}$ and $\mu$
is a chemical potential governing the number of occupied vertices.
Assuming that the graph is a tree, we can write exact equations for
the probability of having configuration $(\sigma_i, \{\sigma_k
| k \in \partial i \setminus j \})$, on node $i$ and its neighbors except for
$j$, when constraints $I_j$, $I_i$ and $I_{ij}$
are absent (cavity graph). We denote this probability $\nu_{i \to
  j}(\sigma_i,\sigma_{i \to j})$ and write
\begin{equation}
\nu_{i \to j}(\sigma_i,\sigma_{i \to j}) = \frac{1}{Z_{i \to j}} \sum_{\sigma_{k \to i}} \prod_{k \in \partial i \setminus j}  I_k I_{ik} \nu_{k \to i}(\sigma_k,\sigma_{k \to i}) e^{-\mu \sigma_i}
\end{equation}
where $\sigma_{i \to j} = \{\sigma_k | k \in \partial i \setminus j \}$ and
$Z_{i \to j}$ is a normalization constant.
The equations, called Belief Propagation (BP) equations \cite{YFW03}, are exact on
tree graphs, but can be used on general graphs to find an
estimate of the above marginals. They give a good approximation if the
graph is locally tree-like, i.e. there are only large loops whose
length diverges with the system's size, as for random graphs of finite
average degree.
Beside this, in writing the BP equations we assume that only
one Gibbs state describes the system, i.e. we have Replica Symmetry.

For the present problem, the BP equations simplify considerably if we
write them in terms of variables $R_{\sigma_i,m}^{i \to j}
\equiv \nu_{i \to j}(\sigma_{i}, m)$ in which $m$ is the number of
occupied neighboring nodes of $i$ (in the cavity graph). Looking at the BP equations
one realizes that a configuration satisfying all constraints
(i.e. solution of the BP equations) contains only three
kinds of these variables: 1) the probability that a node is
occupied in the cavity graph and all its neighbors are
empty $r_1^{i \to j}=R_{1,0}^{i \to j}$, 2) the probability that a
node is empty as well as all its neighbors $r_{00}^{i \to
  j}=R_{0,0}^{i \to j}$, and 3) the probability that a node is empty
but not all neighbors are empty $r_0^{i \to j}=\sum_{m=1}^{k_i-1}R_{0,m}^{i \to j}$.
In terms of these variables, the RS cavity equations become
\begin{eqnarray}\label{BPeq1}
r_1^{i \to j} & \propto & e^{-\mu} \prod_{k \in \partial i \setminus j} (1- r_1^{k
  \to i}), \\ \nonumber
r_0^{i \to j} & \propto & \prod_{k \in \partial i \setminus j} (1 - r_{00}^{k \to
  i}) - \prod_{k \in \partial i \setminus j} r_{0}^{k \to i} \\ \nonumber
r_{00}^{i \to j} & \propto & \prod_{k \in \partial i \setminus j} r_{0}^{k \to i}
\end{eqnarray}
With the correct normalization factor, these equations can be
solved by iteration on a given graph.

\begin{figure}
\includegraphics[width=10cm,angle=0]{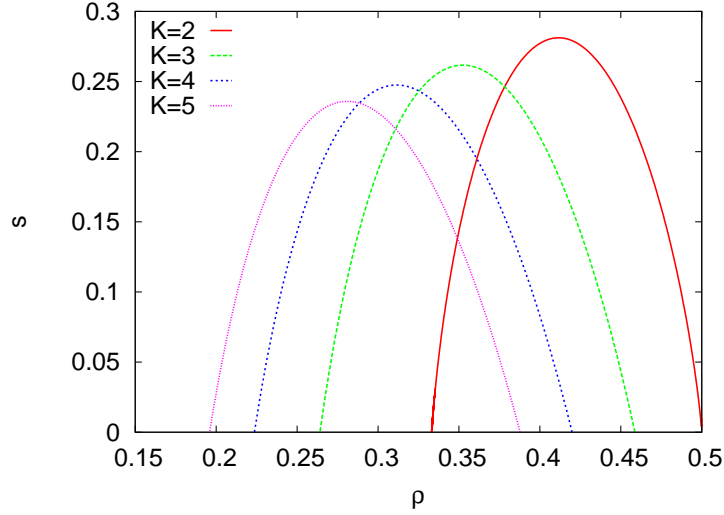}
\caption{(Color online) BP entropy $s(\rho)$ of maximal independent sets for
  random regular graphs with different values of $K=2,3,4,5$ (curves from right to left).}\label{fig_entRRG}
\end{figure}

\subsubsection{Random Regular Graphs}
In random regular graphs with degree $K$, the equations do
not depend on the edge index $i \to j$ at their fixed point, and we have
\begin{eqnarray}\label{BPeq2}
r_1 & = &\frac{e^{-\mu}(1-r_1)^{K-1}}{e^{-\mu}(1-r_1)^{K-1}+(1-r_{00})^{K-1}},\\ \nonumber
r_0 & = &\frac{(1-r_{00})^{K-1}-r_0^{K-1}}{e^{-\mu}(1-r_1)^{K-1}+(1-r_{00})^{K-1}},\\ \nonumber
r_{00} & = & \frac{r_0^{K-1}}{e^{-\mu}(1-r_1)^{K-1}+(1-r_{00})^{K-1}}.
\end{eqnarray}

The zero temperature partition function counts the number of ground
states of the system, i.e. the number of mIS, weighting each occupied
vertex with a factor $e^{-\mu}$. The corresponding free energy is defined as
\begin{equation}
e^{-\mu N f(\mu)} = Z = \int d\rho e^{N s(\rho) - \mu N \rho}
\end{equation}
in which we have decomposed the sum over mIS configurations in
surfaces at the same density of occupied sites $\rho$, isolating the entropic
contribution at each density value $s(\rho)$. The knowledge of the
free energy $f(\mu)$ allows to compute by Legendre transform the
behavior $s(\rho)$ of the entropy of mIS as a function of the density
of occupied vertices, that can be compared with the results obtained
by means of the annealed calculation (Appendix \ref{appANN}).
In the Bethe approximation, the free energy can be computed as
\begin{equation}
\mu f=\frac{1}{N} \sum_i \mu\Delta f_i-\frac{1}{N} \sum_{(i,j)\in \mathcal{E}} \mu\Delta f_{ij},
\end{equation}
where
\begin{eqnarray}
e^{-\mu  \Delta f_i}  & =  & \sum_{\sigma_i,\sigma_{\partial
    i},\sigma_{k \in \partial j \setminus i}} I_{i} \prod_{j \in \partial i}
\nu_{j \to i}(\sigma_j, \sigma_{j \to i})    \\
e^{-\mu  \Delta f_{ij}} & = & \sum_{\sigma_i,\sigma_j,\sigma_{\partial
    i \setminus j},\sigma_{\partial j \setminus i}} I_{ij} \nu_{i \to j}(\sigma_i,
\sigma_{i \to j}) \nu_{j \to i}(\sigma_j, \sigma_{j \to i})
\end{eqnarray}
and in terms of the variables $\{r_1, r_0, r_{00} \}$
\begin{eqnarray}
e^{-\mu  \Delta f_i}  & =  & e^{-\mu}(1-r_1)^K+(1-r_{00})^K-r_0^K \\
e^{-\mu  \Delta f_{ij}}  & = &  r_0^2+2r_1(r_0+r_{00})
\end{eqnarray}
Moreover in these variables the density is easily written as
\begin{equation}
\rho=\frac{e^{-\mu}(1-r_1)^{K}}{e^{-\mu}(1-r_1)^{K}+(1-r_{00})^K-r_0^K}.
\end{equation}
Once we have solved the BP equations (\ref{BPeq2}), we have $\rho$ and
$f(\mu)$ and the Legendre transform
$\mu f(\mu) = -\max_{\rho} \left[ s(\rho) - \mu \rho  \right]$, so
we can compute the entropy $s(\rho)$ by inverting it.

We have solved the BP equations numerically and plotted the corresponding $s$ vs. $\rho$ diagrams in Fig.\ref{fig_entRRG} for several values of the degree $K = 2,3,4,5$.

At this point one should check the stability of BP equations at the
fixed point. This stability is important to have convergence and find
reliable values for the marginals. Notice that BP stability is not a
sufficient condition for the problem to be in the RS phase, but
a necessary one, i.e. if BP are unstable, then we can
conclude that RS assumption is not correct anymore \cite{RBMM04}.
In general, two kinds of instabilities can occur: a {\em ferromagnetic
instability} (or modulation instability), associated with the divergence
of the linear susceptibility and signaling the presence of a continuous
transition toward an ordered state; and a {\em spin-glass instability},
associated with the divergence of the spin-glass susceptibility and
signaling the existence of RSB and possibly a continuous spin-glass
transition.
In the present case, since we are dealing with models defined on
random graphs, the first kind of instability is excluded and we focus
on the latter one.

\begin{figure}
\includegraphics[width=10cm,angle=0]{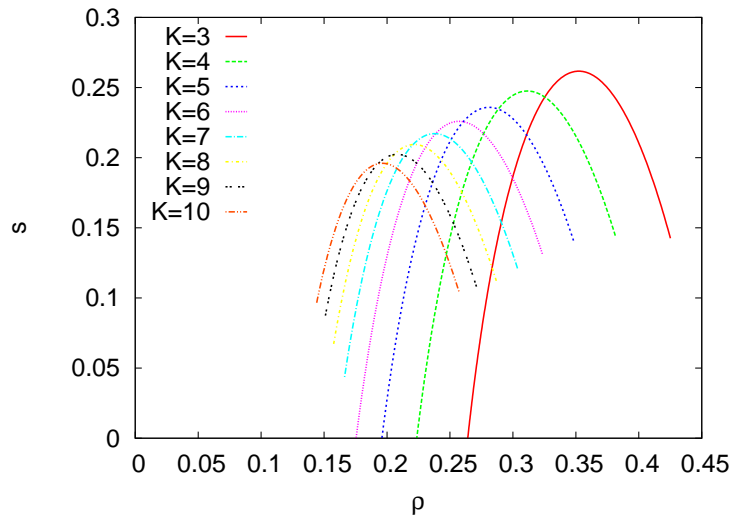}
\caption{(Color online) RS entropy $s(\rho)$ for random regular graphs of
  various degree $K = 3-10$ (curves from right to left) and size $N=10^4$. The curves are plotted only in the
  intervals of density in which BP equations converge.}\label{fig_entropy-stability}
\end{figure}

The details of stability analysis are reported in Appendix
\ref{appSA}. Called $\lambda_M$ the maximum eigenvalue of the
stability matrix $M$, the stability condition is $\Lambda
\equiv (K-1) \lambda_M^2 <1$. In random regular graphs this happens as
long as $\rho < \rho_{s_2}^{BP}$. For larger densities the BP
equations do not converge even if we use a linear combination of old
and new messages to stabilize the dynamics (like in learning processes).
On the other side, for $\rho < \rho_{s_1}^{BP}$ we have $\Lambda <1$
but for $K > K_{m} = 6$ the algorithm converges only if we stabilize it. The numerical
values for different $K$ are given in Table \ref{table2}.

In Fig. \ref{fig_entropy-stability} we plotted again the curves of the RS entropy as function of $\rho$ for various degree values, now showing only the region of the curve in which the RS solutions are stable.
It is worthy noting that for $K \leq K_{m} = 6$ BP equations converge very easily in the low density region while they are always unstable for $K > K_{M}=2$ in the high density region.

\subsubsection{Erd\"os-R\'enyi Random Graphs}

\begin{figure}
\includegraphics[width=10cm,angle=0]{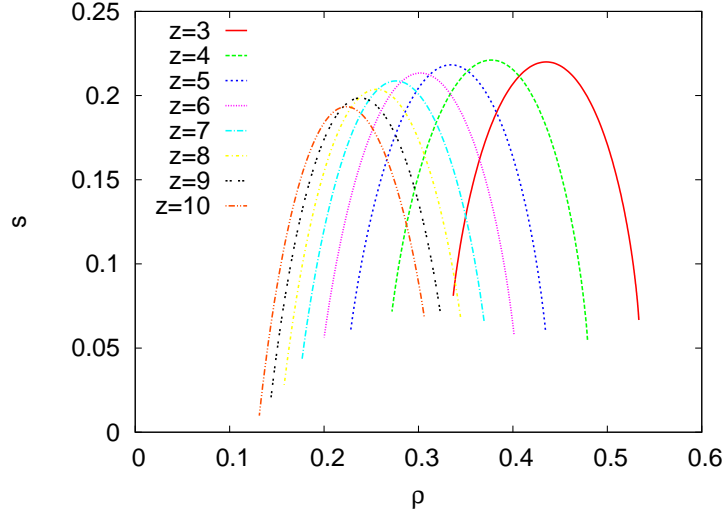}
\caption{(Color online) Entropy of random ER graphs of average degree $z = 3-10$ (curves from right to left) and size
  $N=10^4$ in the range of densities in which BP equations converge.
}\label{fig_entropy-stability-z}
\end{figure}

Consider the ensemble of ER random graphs with degree distribution
$p_k$ and average degree $z$. In this case the probabilities
$\underline{r}\equiv \{r_1, r_0, r_{00}\}$ depend on edge index
$(i \rightarrow j)$:

\begin{eqnarray}
r_1^{i\rightarrow j}\propto e^{-\mu}\prod_{k\in \partial i \setminus j }(1-r_1^{k\rightarrow i}),\\
\nonumber
r_0^{i\rightarrow j}\propto \prod_{k\in \partial i \setminus j }(1-r_{00}^{k\rightarrow i})-\prod_{k\in \partial i \setminus j }r_0^{k\rightarrow i},\\
\nonumber r_{00}^{i\rightarrow j}\propto \prod_{k\in
\partial i \setminus j}r_0^{k\rightarrow i},
\end{eqnarray}
One can run the above equations on random graphs to obtain the BP
entropy.
Figure \ref{fig_entropy-stability-z} displays the results
that we obtain in this way. There are some regions in which the equations do not converge
even if we use a linear combination of new and old messages to
stabilize the equations.

Let us denote the above BP equations by $\mathcal{BP}$. In a general random graph messages $\underline{r}$ change
from one directed edge $(i\rightarrow j)$ to another. In a large graph (or equivalently in the ensemble of random graphs) the statistics of these fluctuations is described by
$P(\underline{r})$ which satisfies
\begin{eqnarray}
P(\underline{r})=\sum_k q_k \int \prod_{l=1}^k dP(\underline{r}^l)
\delta(\underline{r}-\mathcal{BP}),
\end{eqnarray}\label{ERRSPr}
where $q_k=(k+1)p_{k+1}/z$ is the excess degree distribution.

\begin{figure}
\includegraphics[width=10cm,angle=0]{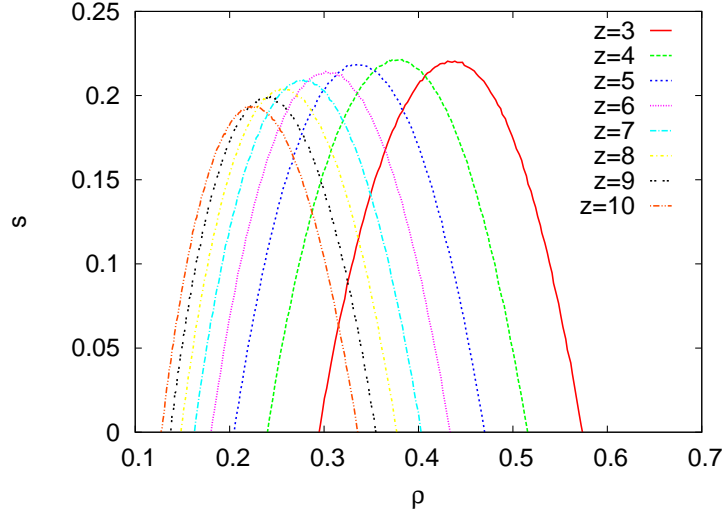}
\caption{(Color online) Entropy of random ER graphs of average degree $z=3-10$ (curves from right to left) obtained with population dynamics.
}\label{entropy-pop}
\end{figure}

To obtain the entropy, we have to solve the above equation by means of
population dynamics (see App. \ref{appPOP}). This is a well known method to solve equations involving
probability distributions \cite{MP02}.  Having $P(\underline{r})$ we compute the free energy in the Bethe approximations
$\mu f = \mu \langle \Delta f_i \rangle -\frac{K}{2} \mu \langle \Delta f_{ij} \rangle$, with
\begin{eqnarray}
\nonumber -\mu \langle \Delta f_i \rangle & = &\sum_k p_k \int \prod_{l=1}^k
dP(\underline{r}^l) \ln \left( e^{-\mu}\prod_{l}(1-r_1^l)+\prod_{l}(1-r_{00}^l)-\prod_{l}r_0^l \right),\\
\nonumber -\mu \langle \Delta f_{ij} \rangle &= &\int
dP(\underline{r})dP(\underline{r}')\ln \left(r_0r_0'+r_1(r_0'+r_{00}')+r_1'(r_0+r_{00}) \right),
\end{eqnarray}
and finally invert the Legendre transform.
The density of covered nodes is given by
\begin{eqnarray}
\rho & = &\sum_k p_k \int \prod_{l=1}^k
dP(\underline{r}^l)\frac{e^{-\mu}\prod_{l=1}^k(1-r_1^l)}{e^{-\mu}\prod_{l=1}^k(1-r_1^l)+
\prod_{l=1}^k(1-r_{00}^l)-\prod_{l=1}^k r_0^l}.
\end{eqnarray}
Notice that in this way we obtain the entropy in the RS approximation for infinite random graphs.
The numerical values are already in good agreement with those obtained on single samples of size $N=10^4$.
It should be mentioned that in the population dynamics algorithm we do not have the convergence
problem, we could always stabilize the dynamics and find the entropy in the whole region
of densities. Figure \ref{entropy-pop} displays the results of population dynamics for the entropy.
The lower and upper bounds in Table \ref{table1} have been obtained with population dynamics.

\subsection{Replica Symmetry Breaking}

Here we go beyond the RS ansatz and explore the
possibility that maximal independent sets organize in clusters or in
even more complex structures \cite{MPV87}.
The instability of BP equations that we observed
in the previous section could be a clue of a transition into a 1RSB or
full RSB (fRSB) phase.
This would be reasonable at least for the high density
region because the maximum mIS (MIS) coincides with the minimum vertex
covering for the conjugate problem, that is presumed to present RSB of
order higher than 1 (possibly fRSB) \cite{WH00,Z05,BH04}.

In order to verify these ideas we consider 1RSB solutions of the
mIS problem.

Having a 1RSB phase means that the solution space is composed of well separated (distances of order $N$) clusters of solutions. Each cluster has its
own free energy density $f_c$ and there are an exponential number of clusters of given free energy $e^{N\Sigma(f)}$ where
$\Sigma(f)$ is called complexity.
The physics of this phase is described by the following generalized partition function:

\begin{eqnarray}
\mathcal{Z}=e^{-m\mu N \Phi}=\sum_c e^{-m\mu f_c}= \int df e^{N[\Sigma(f)- m\mu f]},
\end{eqnarray}

Here $m$ is the Parisi parameter that describes 1RSB phase. Having the distribution of cavity fields $\mathcal{P}(\underline{r}^{i \rightarrow j})$
among the clusters we write the following expression for $\Phi$ in the Bethe approximation \cite{MPR05}:

\begin{eqnarray}
y N \Phi= \sum_i y \Delta \Phi_i-\sum_{(i,j)\in \mathcal{E}} y\Delta \Phi_{ij},
\end{eqnarray}

where $y=m\mu$ and

\begin{eqnarray}
e^{ -y \Delta \Phi_i}=\int  \prod_{j \in i}d \mathcal{P}(\underline{r}^{j \rightarrow i}) e^{-y \Delta f_i},\\ \nonumber
e^{ -y \Delta \Phi_{ij}}=\int  d\mathcal{P}(\underline{r}^{i \rightarrow j})d\mathcal{P}(\underline{r}^{j \rightarrow i}) e^{-y \Delta f_{ij}}.
\end{eqnarray}

The cavity fields distribution satisfies

\begin{eqnarray}
\mathcal{P}(\underline{r}^{i \rightarrow j})\propto \int  \prod_{k \in \partial i \setminus j}d \mathcal{P}(\underline{r}^{k \rightarrow i}) e^{-y \Delta f_{k\rightarrow i}} \delta(\underline{r}^{i \rightarrow j}-\mathcal{BP}).
\end{eqnarray}\label{1RSBPr}

where we have assumed that the graph is regular and so $\mathcal{P}(\underline{r}^{i \rightarrow j})$
does not depend on the edge label. We have also introduced the cavity free energy change

\begin{eqnarray}
e^{-\mu\Delta f_{i\rightarrow j}}=e^{-\mu}(1-r_1)^{K-1}+(1-r_{00})^{K-1}.
\end{eqnarray}

From the generalized free energy $\Phi$ we can obtain the complexity by a Legendre transform

\begin{eqnarray}
\Sigma(f)= -y\Phi+ y f, \hskip1cm f=\frac{\partial y\Phi}{\partial
y}.
\end{eqnarray}\label{Sigmaf}

A simplifying approach would be that of  working in
the limit of Survey Propagation (SP) \cite{MeZ02}, assuming infinite chemical potential $\mu \to \pm \infty$ and zero Parisi parameter $m \to
0$, with finite ratio $y = m \mu$. This means that we focus only on the most
numerous clusters ($m=0$) composed of frozen solutions
($\mu \to \pm \infty$). However, this is not consistent with the spatial organization of mISs emerging from Section \ref{orgSEC}.
Indeed, we know that variables are locally but not globally frozen,
 and the absence of globally frozen variables means that if there is any cluster of solutions it contains only unfrozen variables \cite{S08}.
Therefore we do not expect to find any physically relevant result by means of Survey Propagation.
In Appendix \ref{appSP}, we report a detailed analysis showing that SP complexity is indeed unphysical.

Let us relax the $m=0$ restriction to see if there is any other
type of clusters. Notice that when $m$ is finite the chemical
potential can also be finite and can not have frozen variables in the clusters (i.e. variables assuming always the same value for all solutions in one cluster).
A nonzero complexity in this case would imply the presence
of unfrozen clusters.

For a value of $m$ the relevant clusters are those that maximize
$\Sigma(f)-m\mu f$. The parameter $m$ itself is chosen to maximize the
free energy $\Phi(m)$. As long as we are in the RS phase, $m^*=1$
clusters are the thermodynamically relevant ones with zero
complexity. A dynamical transition could occur if the complexity
of these clusters $\Sigma(m=1)$ takes a nonzero value. But a real
thermodynamic transition (1RSB) occurs when the relevant clusters
have again a zero complexity with $m^*<1$.

\begin{figure}
\includegraphics[width=10cm,angle=0]{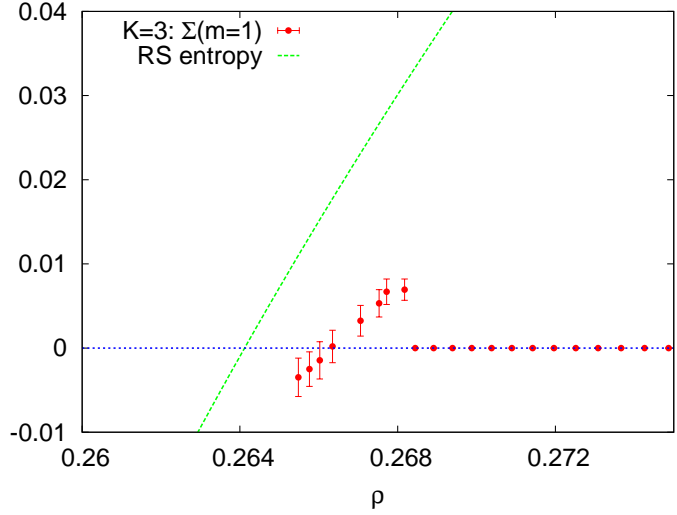}
\caption{(Color online) Comparing RS entropy (dashed line) and $\Sigma(m=1)$ (circles) in the low density region.}\label{cov-1RSB}
\end{figure}

Here we resort again to population dynamics (see App.
\ref{appPOP}) to solve Eq. \ref{1RSBPr} and to find the complexity
as described above. For simplicity we only consider the case of
random regular graphs with $K=3$.  We have used a population of size $N_p=10^5$ to represent the distributions. To get the
nontrivial fixed point of the dynamics we start from completely polarized messages \cite{MMo06} and update the population
for $T=10^4$ iterations to reach the equilibrium. In each iteration all members of the population are updated in a random sequential way.
After this equilibration stage we take $M=10^3$ independent samples of the population to compute the free energy $\Phi$ and other
interesting quantities like the complexity.

\begin{figure}
\includegraphics[width=10cm,angle=0]{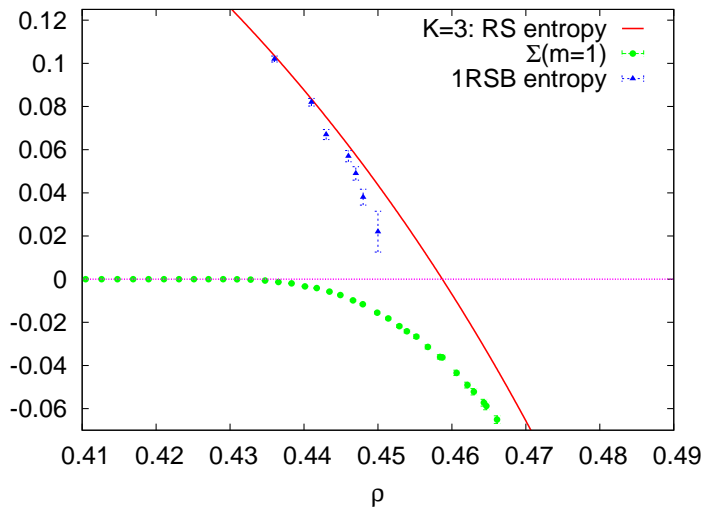}
\caption{(Color online) Comparing RS entropy (solid line), 1RSB entropy (triangles) and $\Sigma(m=1)$ (circles) in the high density region.}\label{pac-1RSB}
\end{figure}

In Figs. \ref{cov-1RSB} and \ref{pac-1RSB} we report the $m=1$ complexity
obtained with population dynamics and compare it with the RS entropy in the extreme density regions.

There is a small interval in the low density region in which $\Sigma(m=1)$ is positive, signaling a dynamical
transition in that region. Since the RS solution is stable in this region, total entropy will be equal to the RS one
and the minimum density is the point that $m=1$ complexity vanishes.
Despite the large equilibration time we have still large statistical errors
maybe due to the instability of this solution. We observed that the errors improve very slowly by increasing size of population or number of samplings, indicating
a poor convergence of this solution.

In the high density region we observe a completely different behavior where $\Sigma(m=1)$ is always non-positive.
This behaviour is also observed in other problems like $3$-SAT \cite{MRS08}, and it means that in the high density region
dynamical and condensation transitions coincide \cite{KMRSZ07}. Below the transition, complexity is zero and we are in the RS phase with $m^*=1$; for larger densities, $\Sigma(m=1)$ becomes
negative and we have a condensation transition where $m^*<1$. The entropy computed at this value of $m$ gives the 1RSB prediction of the entropy.
The maximum density that we obtain in this way is displayed in Fig. \ref{pac-1RSB}. We recall that a maximum mIS is complement of a minimum vertex covering
and we already know that these coverings have a nonzero entropy \cite{Z05,WH00}.

We found qualitatively the same behavior for random regular graphs of degree $K=4,5$.

\subsection{Distance from a solution}

In Section \ref{orgSEC}, we have proved that in a general graph of size $N$ a solution (mIS) is at a finite
distance $\mathcal{O}(\langle k^2 \rangle)$ from a number of
$\mathcal{O}(N)$ other solutions (mIS). The mathematical proof is based on the idea that by flipping a
single variable in a mIS configuration we generate a rearrangement
process that propagates at most to the second neighbors of the flipped variable.
Here we substantiate this result by means of a statistical mechanics
calculation.

\begin{figure}
\includegraphics[width=10cm,angle=0]{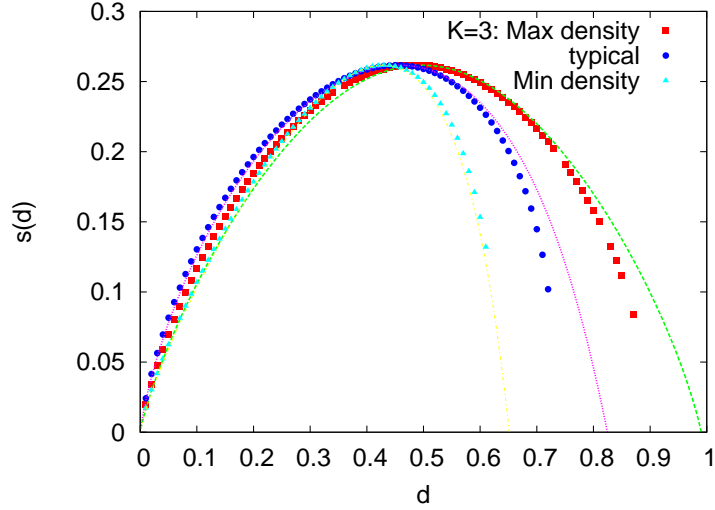}
\caption{(Color online) Comparing analytic and numerical results for $s(d)$ in random regular graphs of degree $K=3$ and size $N=10^4$. The points have been obtained by computing
$s(d)$ for a given solution using the cavity method. Max density (squares) and Min density (triangles) refer to
the best extreme solutions that we obtain.}\label{fig_distance}
\end{figure}

In the large deviations cavity formalism, it is possible to compute
the number of solutions of a CSP at a distance $d$ from
a given one $\underline{\sigma}^*$ using the weight enumerator function \cite{MPR05,MM06,DRZ08}
\begin{equation}
Z=e^{-Nx f^*} = \sum_{\underline{\sigma}} \prod_i I_i
\prod_{(i,j)\in \mathcal{E}} I_{ij} e^{-x \sum_i
(\sigma_i-\sigma_i^*)^2}=\int_d e^{N[s^*(d)-x d]}.
\end{equation}
where $s(d)$ is the entropy of solutions at a distance $d = \frac{1}{N}\sum_i (\sigma_i - \sigma_i^*)^2$ from $\underline{\sigma}^*$.
Averaging over all solutions (mIS)
\begin{equation}
\overline{Z}=e^{-Nx f}=\frac{1}{Z_0}\sum_{\underline{\sigma}^*}
\sum_{\underline{\sigma}} \prod_i I_i \prod_{(i,j)\in \mathcal{E}}
I_{ij}  e^{-x \sum_i
  (\sigma_i-\sigma_i^*)^2 - \mu \sum_i \sigma_i^*}=\int_d e^{N[s(d)-x d]},
\end{equation}
where we added a chemical potential $\mu$ to keep track of the
contribution of mISs with different density. Notice that here we are using
the annealed approximation which works if there are no strong fluctuations in $Z$.
We find $f$ in the Bethe approximation in terms of cavity fields (now weighted with the term $e^{-x(\sigma_i-\sigma_i^{*})^2}$) and extract the expression $s(d)$ of the entropy of solutions at an Hamming distance $dN$ from another
by means of Legendre transform,
\begin{eqnarray}
-x f & = & \max_d [s(d)-x d], \hskip1cm d=f+x\frac{\partial f}{\partial x}.
\end{eqnarray}
For the mIS problem, the cavity equations have now to take into
account the current value of the variable $\sigma_{i}$ and that in the
reference configuration $\sigma_i^*$,
\begin{equation}
\nu_{i \to j}(\sigma_i,\sigma_{i \to j}| \sigma_i^*) \propto
e^{-x(\sigma_i-\sigma_i^*)^2 - \mu \sigma_i^*} \sum_{\sigma_{k \to i}} \prod_{k
  \in \partial i \setminus j}  I_k I_{ik} \nu_{k \to i}(\sigma_k,\sigma_{k \to
  i}| \sigma_k^*)
\end{equation}

Knowing $f$ and $d$ we can numerically compute $s(d)$, as reported in
Fig.\ref{fig_distance} for the case $K = 3$. The plot shows that at typical densities there is always an extensive
number of mISs at any finite distance $d$ from another mIS. The same
holds for non-typical values of the density (obtained with $\mu \neq
0$), even if in this case we can push our analyses only up to those
values at which BP converges. From the figure we observe that a low density solution
is closer to other solutions than a high density one whereas typical solutions lie in between.

\section{Numerical Simulations}
\label{numSEC}
In this section we discuss different classes of numerical simulations that can be used to investigate the properties of maximal independent sets on random graphs. We first consider some greedy algorithms, that generate a mIS in a time that scales linearly with the system's size.
These algorithms work in a very limited range of density values, though they are of interest for their application to the study of the best-response dynamics in strategic network games \cite{BK07,GGJVY08}. \\
A more effective way to explore non typical regions of the phase diagram, at low and high densities, is by means of Monte Carlo simulations based on the lattice gas representation.
Monte Carlo methods are also useful to obtain an estimate of the entropy of maximal-independent sets by thermodynamic integration.  Other interesting results can be obtained by means of a particular kind of zero-temperature Monte Carlo simulation, that allows transitions from a mIS directly to another and have been conceived appositely for sampling the space of maximal-independent sets.\\
Finally, we compare the results of Monte Carlo simulations with those obtained by a completely different technique, the numerical decimation method based on belief-propagation equations (BP decimation, BPD). Both Monte Carlo and BP decimation are effective in sampling mISs in the regions of low and large density values, but they are not able to reach the extreme density limits predicted by theoretical calculations (RS bounds).

Note that the use of a simulated annealing scheme makes the computational time of all MC simulations much longer than that of BP-based algorithms. Therefore, whereas in the case of BP decimation we  present results for systems of $N=10^4$ nodes,  all MC results will be restricted to smaller size ($N = 10^3$) in order to have a reasonable statistics in particular at non-typical densities.

\subsection{Greedy algorithms and Best Response dynamics}

The simplest algorithm to generate maximal independent sets is the {\em Gazmuri's algorithm} \cite{G94},
\begin{itemize}
\item[0.] Start assuming all nodes of the graph to be $0$;
\item[1.] At each time step select a node $i$ uniformly at random, assign $1$ to the node $i$ and remove it from the graph together with all its neighbors (that are $0$ valued) and all the edges departing from these nodes.
\item[2.] Repeat point $1.$ until the graph is empty.
\end{itemize}
The configuration of the removed nodes defines a maximal independent set for the original graph.
The Gazmuri's algorithm works in linear time on every graph, but does not allow any control on the density of covered nodes; therefore one could expect to obtain on average solutions of {\em typical} density, close to the maximum of the entropic curve $s(\rho)$ in Fig. \ref{fig_entropy-stability}-\ref{fig_entropy-stability-z}.

\begin{figure}
\includegraphics[width=8cm,angle=0]{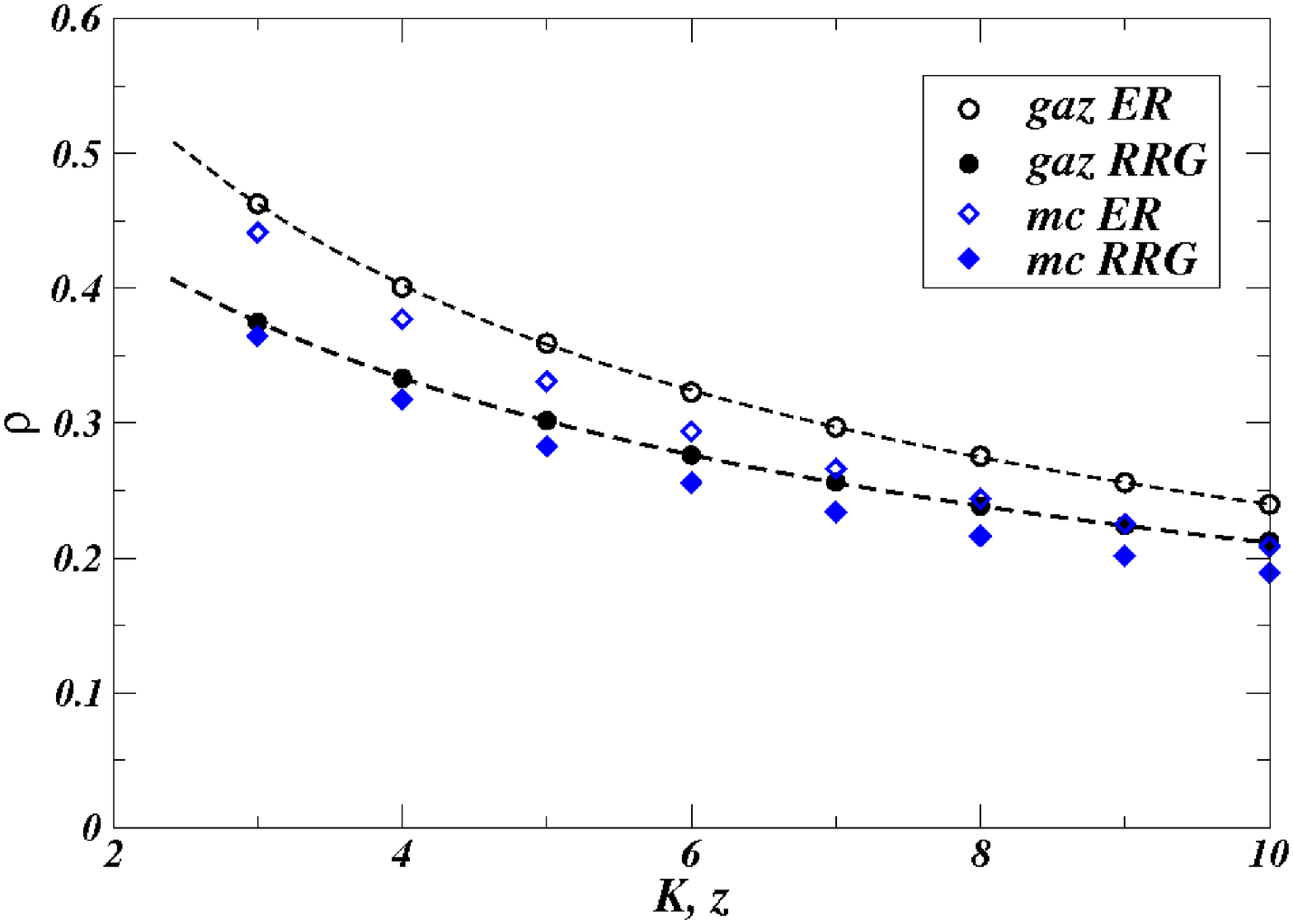}
\caption{(Color online) Average density of coverings $\rho$ (circles) vs. average
  degree $K, z$ in the mISs obtained with Gazmuri's algorithm on ER
  random graphs (open symbols) and random regular graphs (full
  symbols). Data points were obtained averaging over $100$ graphs of
  $N=10^3$ nodes. The dashed lines are the theoretical predictions
  obtained solving the corresponding differential equations for the
  concentration of covered nodes. The typical behavior of greedy
  algorithms is compared with typical results of Monte Carlo
  simulations on the same graphs (diamonds).
} \label{ga1}
\end{figure}

\begin{figure}
\includegraphics[width=10cm,angle=0]{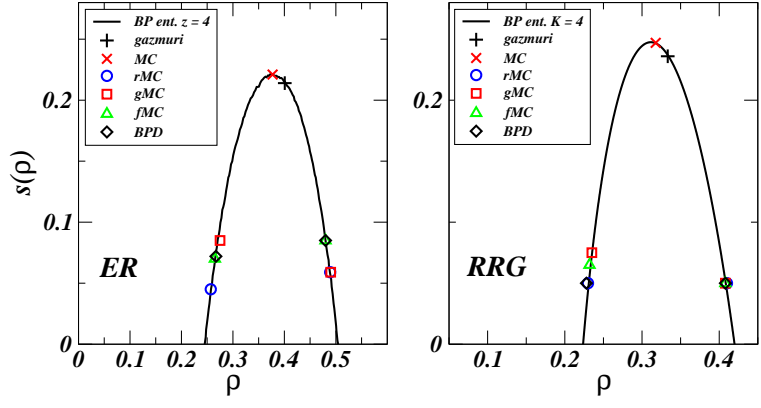}
\caption{(Color online) Entropy curves $s(\rho)$ obtained with BP on both ER (left)
  and random regular graphs (right). On the curves we report the
  minimum and maximum densities at which we can find mIS using
  different algorithms: the Gazmuri's algorithm (black plus symbols),
  standard Monte Carlo (red crosses), rearrangement Monte Carlo (blue
  circles), Monte-Carlo with chemical potential fixing the average
  $\rho$ (red squares), fixed density Monte Carlo (green triangles), BP decimation (black diamonds).
  Data are obtained for graphs with $N=10^3$ nodes and average degree $z,K = 4$.
} \label{mc4}
\end{figure}

ER random graphs are particularly simple to  study theoretically. We assume that the degree distribution remains poissonian during nodes removal, but with a time-dependent average degree.
This is reasonably correct if the random removal is an uncorrelated process. \\
The dynamics of the Gazmuri's algorithm can be expressed in terms of differential equations for concentrations, using a standard mean-field approach recently formalized in probability theory by Wormald \cite{W95}.
The initial number of nodes in the original graph is $N$, but at each step of the process one node and all its neighbors are removed. If we call $c(T)$ the average degree of a node in the graph after $T$ temporal steps, the expected variation of the number of nodes in the graph  is  $\mathbb{E} [N(T+1)-N(T)] = -1 - c(T)$. Writing $N(T)=N n(T/N)$, we get
 the concentration law $\dot{n}(t) = -c(t) -1$. Note that the average degree evolves as $c(t) = c(0) n(t)$ with $c(0)=z$. The two equations give $c(0) n(t) = (c(0) + 1) e^{-c(0) t} - 1$, that vanishes at $t_{f} = \log{(c(0)+1)}/c(0)$. As we cover only one node per unit of time, $\rho(t_f)=t_f$. The Gazmuri's algorithm on ER random graphs of average degree $z$ produces mIS of typical density $\rho_{gaz} \equiv \rho(t_f) = \log{(z+1)}/z$.
In Figure \ref{ga1} we show that this result is in good agreement with
simulations done averaging over $100$  ER graphs of size $N=10^3$ and
corresponds approximately to the typical density, even if it
systematically overestimates the values maximizing the RS entropy (see also Fig. \ref{mc4}).

\begin{figure}
\includegraphics[width=10cm,angle=0]{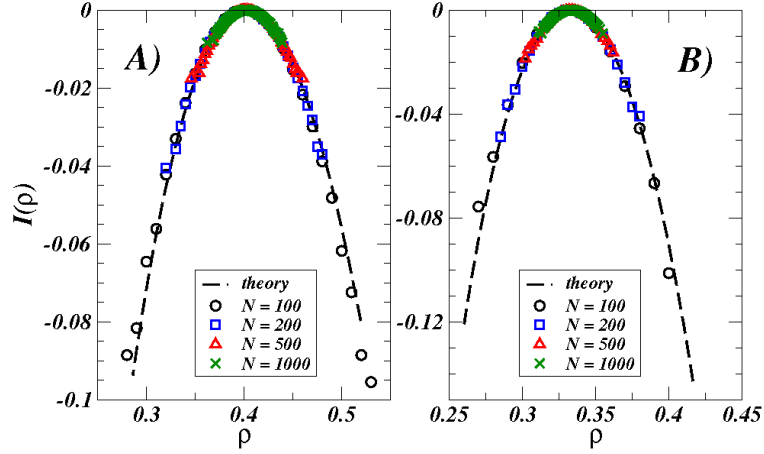}
\caption{(Color online) Rare events in the Gazmuri algorithm for ER random graphs
  (left)  and random regular graphs (right) with $z, K=4$. The
  theoretical behavior of the large deviation functional
  $\mathcal{I}(\rho)$ (dashed line) is compared with results of
  simulations for graphs with $N=100$ (circles), $200$ (squares),
  $500$ (triangles), $1000$ (crosses) nodes. Data points are obtained
  computing the probability $P(\rho)$ of observing mIS of density
  $\rho$ out of  $10^5$ trials. Plotting the rescaled function $\log
  P(\rho) /N$, we find perfect agreement with the theoretical values for $\mathcal{I}(\rho)$.
} \label{ga2}
\end{figure}

The probability of obtaining a non typical mIS with the Gazmuri's algorithm decays exponentially like $e^{N \delta \rho}$ where $\delta \rho$ is the deviation of the density of covered nodes with respect to the typical density $\rho_{gaz}$.
In principle, repeating an exponential number of times the Gazmuri's algorithm, we have non zero probability to find rare trajectories in which
the final density of covered nodes may differ considerably from the
typical one. These finite size effects can be quantified using the following path-integral approach \cite{MZ02,HWbook}.\\
The evolution of the algorithm is fully specified by the evolution of the concentration of covered nodes $x(t)$, the concentration of the number of untouched nodes $n(t)$ or equivalently the evolution of the average degree $c(t)$.  The probability that in the $T+1 ^{th}$ temporal step of the algorithm the number of covered nodes and untouched nodes changes respectively of $\Delta X$ and $\Delta N$ is
\begin{equation}
P^{T+1}_{T}(\Delta X, \Delta N) = e^{-c(t)} \left[ \delta_{\Delta X,1} \sum_{k=0}^{\infty} \delta_{\Delta N,-k-1} \frac{c(t)^k}{k!} \right]
\end{equation}
where we have used the Poisson degree distribution $p_k = \frac{c(t)^k}{k!}e^{-c(t)}$. In the Fourier space
\begin{eqnarray}
\hat{P}^{T+1}_{T}(\xi(t), \mu(t)) & = &\sum_{\Delta X = -\infty}^{\infty}  \sum_{\Delta N = -\infty}^{\infty} P^{T+1}_{T}(\Delta X, \Delta N)  e^{-i \mu(t) \Delta N - i \xi(t) \Delta X} \\
& = & \exp{\left[ -c(t) + i \mu(t) - i \xi(t) + c(t) e^{i\mu(t)} \right]} .
\end{eqnarray}
Then considering $N\Delta T$ consecutive steps and neglecting subleading terms in $\Delta t$ we get
\begin{equation}
P^{T+\Delta T}_{T}(\Delta X, \Delta N) = \int_{-\pi}^{\pi} \frac{d\xi(t)}{2\pi}  \int_{-\pi}^{\pi} \frac{d\mu(t)}{2\pi}  e^{i \mu(t) \Delta N + i \xi(t) \Delta X}  \exp{\left\{ N \Delta t \left( -c(t) + i \mu(t) - i \xi(t) + c(t) e^{i\mu(t)} \right)\right\}}
\end{equation}
where one should then use $\Delta X \simeq N \dot{x}(t) \Delta t$ and $\Delta N \simeq N \dot{c}(t) \Delta t / c(0)$.
The probability $P(x_f|c)$ of a trajectory $\{x(t),c(t)\}$ given the initial and final conditions $\{x(0)=0,c(0)=z\}$ and $\{x(t_f)=x_f=t_f,c(t_f)=0\}$ is
\begin{equation}
P(x|c) = \prod_{t<1} \int_{-\pi}^{\pi} \frac{d\xi(t)}{2\pi}  \int_{-\pi}^{\pi} \frac{d\mu(t)}{2\pi} \int_0^1 dx(t) \int_0^c dc(t) \exp{\left\{ - N \Delta t \mathcal{L}(\dot{x}(t),\dot{c}(t),x(t),c(t),\xi(t),\mu(t)) \right\}}
\end{equation}
 with Lagrangian
\begin{equation}
\mathcal{L}(\dot{x},\dot{c},x,c,\xi,\mu) = c(t)- i \mu(t)+i\xi(t)-ce^{i\mu(t)} - i\xi(t) \dot{x}(t)- i \mu(t) \frac{\dot{c}}{z}
\end{equation}
The Euler-Lagrange equations are
\begin{eqnarray}
\nonumber \dot{x} & = &1, \\
\nonumber \dot{\xi} & =& 0 , \\
\nonumber \dot{c} & =& -z (1+ ce^{i\mu}) \\
i\dot{\mu} & =& z ( e^{i\mu} -1) \label{ELeq1}
\end{eqnarray}
Solving the equations, the probability of rare events becomes $P(x|c) \approx \exp \left(N \mathcal{I}(x,c) \right)$ with the large deviation functional given by the saddle-point action $\mathcal{I}(x,c) = \int_0^{t_f} dt \mathcal{L}(x,c)$. We have solved numerically the Eqs. \ref{ELeq1} and computed the large deviation functional $\mathcal{I}(\rho,z)$ for ER graphs with different values of the average degree $z$. In Fig. \ref{ga2}-A we have plotted the behavior of the large deviation functional for $z=4$ as a function of the density $\rho$ of covered nodes in the mIS generated by the algorithm.
Its theoretical behavior is compared to the results of simulations of Gazmuri's algorithm on a ER graph with $N=100, 200, 500, 1000$ nodes and same average degree $K=4$. The results are in perfect agreement.

In the case of random regular graphs, it is possible to obtain an analytical estimate of the
behavior of the Gazmuri's algorithm using an approach based on random pairing processes \cite{W99}. A random pairing process is used to generate random graphs with a given degree distribution and consists in taking a number of copies of the same node equal to its degree and in matching these copies randomly with copies of other nodes until the network is formed and no copy remains unmatched.
The evolution of mIS can be described using a random pairing process. We consider two quantities: the number of covered nodes (i.e. that is equal to the time $T$) and
the number of nodes that still have not been touched by the algorithm, i.e. $Y(T)$. All copies of the nodes are initially {\em untouched}: at each step,  we select randomly an untouched copy and cover her together with all her siblings. The $K$ other copies matched with these ones are removed ({\em exposed} in Ref. \cite{W99}). The result is that at each step the number of untouched copies
decreases of $2K$, but some of the remaining copies are siblings of exposed ones. This is important in order to compute the total variation of the number of untouched nodes during the process. In fact, the probability that the pairing of a selected copy is also untouched is given by the number of untouched copy $K Y(T)$ divided by the total number of non-exposed  copies $K N-2 K T$.
This random pairing algorithm is repeated until there are no untouched nodes anymore.
As we are considering the pairing on the fly, on an annealed network structure, we can neglect the evolution of the degree distribution and write an equation for $Y(t)$. Its variation in a single time step is $\mathbf{E}[Y(T+1)-Y(T)] = -1 - Y(t)/(N-2T)$.
The dynamics of $y(t)=Y(T/N)/N$ is governed by the differential equation
\begin{equation}
\frac{d y(t)}{dt} = -1 - K \frac{y(t)}{1-2t}
\end{equation}
 that gives $y(t) = \frac{(K-1)(1-2t)^{K/2}-(1-2t)}{K-2}$. The density of covered nodes is given by the time $t_f$ at which $y(t_f)=0$, i.e.
 $t_f = \frac{1}{2}-\frac{1}{2}(K-1)^{2/(2-K)}$.
Figure \ref{ga1} (full simbols) compares the theoretical prediction
for various values of $K$ with the average density observed in the
simulation of the Gazmuri's algorithm on random regular graphs of size $N=1000$. \\
In a small network, a single realization of the Gazmuri's algorithm can deviate considerably from the average behavior, even if the degree distribution is initially regular.
The fluctuations are now associated with the number of possible untouched nodes exposed by the pairing process in a time step.
This can be easily quantified applying the path-integral method to obtain the large deviation functional. The probability that in the $T+1 ^{th}$ temporal step of the algorithm the number of untouched nodes changes of $\Delta Y$ is
\begin{equation}
P^{T+1}_{T}(\Delta  Y) =  \sum_{n=0}^{K} \delta_{\Delta Y,-1-n} \binom{K}{n} \left( \frac{Y(T)}{N-2T} \right)^n \left( 1- \frac{Y(T)}{N-2T} \right)^{K-n} .
\end{equation}
In Fourier space it becomes
\begin{eqnarray}
\hat{P}^{T+1}_{T}(\xi(T)) & = & \sum_{\Delta Y = -\infty}^{+\infty}\sum_{n=0}^{K} \delta_{\Delta Y,-1-n} \binom{K}{n} \left( \frac{Y(T)}{N-2T} \right)^n \left( 1- \frac{Y(T)}{N-2T} \right)^{K-n} e^{- i \xi(T) \Delta Y}\\
& =& e^{i \xi(T)} \left[ 1+ \frac{Y(T)}{N-2 T} (e^{i \xi(T)}-1)\right]^K
\end{eqnarray}
 The corresponding Lagrangian in the continuum limit is
 \begin{equation}
\mathcal{L}(\dot{y},y,\xi) = - i \xi(t)-i\xi(t) \dot{y}(t) - K \log{\left[1 + \frac{y(t)}{1-2t} (e^{i\xi(t) -1)}\right]}
\end{equation}
 Imposing the stationarity, we find the saddle-point equations
 \begin{eqnarray}\label{ELeq2}
i \dot{\xi} & =& \frac{K(e^{i \xi} -1)}{1-2t+y(e^{i\xi}-1)} , \\
\dot{y} & =& -1- \frac{K y e^{i \xi} }{1-2t+y(e^{i\xi}-1)} .
\end{eqnarray}
Solving the equations and computing the large deviation functional
$\mathcal{I}(\rho,K)$ we find the behavior displayed in
Fig. \ref{ga2}-B (dashed line), in which we also plotted the results of
the numerical simulation of the Gazmuri's algorithm on random regular graphs of small sizes. As for ER graphs, the statistics of rare events is extremely well reproduced by our theoretical calculations.

The dynamics of Gazmuri's algorithm is relevant for economic applications, because it reproduces the main features of the best-response
dynamics in Best-Shot strategic games \cite{GGJVY08}.
In the best-response (BR) dynamics, all variables are initially assigned to be $0$ or $1$ with a given probability $p_{in}$. At each time step a node $i$ is randomly selected: if at least one of the neighbors of $i$ is $1$, then the node $i$ is put to $0$, otherwise if all neighbors are $0$, the node is put to $1$. This type of dynamics has been recently studied by Lopez-Pintado \cite{LP07} on uncorrelated random graphs by means of a dynamical mean-field approach. In random regular graphs of degree $K$, the density of covered nodes (contributors) evolves following the equations
\begin{equation}
\label{lopez-lintado}
\frac{d \rho(t)}{dt} = - \rho(t) (1-(1-\rho(t))^K) + (1-\rho(t)) (1-\rho(t))^K
\end{equation}
and converges rapidly to the fixed point $\rho = (1-\rho)^K$. This solution looks different from that obtained by means of the pairing process; however, the actual process is not very different.
In fact, for the nature of the strategic games associated with the mIS problem, a single sweep of BR over the system is sufficient to reach a Nash Equilibrium (i.e. to satisfy all variables, without generating contradictions). Therefore,  Best-Response dynamics behaves like the Gazmuri's algorithm, apart from choice of the initial conditions that could introduce a bias in the density values. Simulating BR dynamics with different initial bias $p_{in}$, we verified that the final density is almost independent of $p_{in}$ and agrees reasonably well with the results of Gazmuri's algorithm (not shown).

\subsection{Monte Carlo methods applied to mIS problem}

\subsubsection{Different types of simulated annealing}
Monte Carlo algorithms for finding maximal independent sets are based on a simulated annealing scheme for the auxiliary binary spin model  in which the energy $E$ of the system corresponds to the number of unsatisfied local constraints (that we have already defined in Section \ref{SECcavity}) \cite{KGV83}.
Starting from the high temperature region (i.e. random configurations of $0$s and $1$s),  we slowly decrease the temperature to zero, with the following Metropolis rule:
1) pick up a node randomly and flip its binary variable;
2) if the energy is decreasing, then accept the move with probability $1$, otherwise accept the move with a probability $e^{-\beta \Delta E}$.

In the absence of a constraint on the density of covered nodes, the
algorithm always finds a solution of typical density. Fig \ref{ga1}
(diamond-like symbols) shows the dependence of the average density of
covered nodes $\rho$ for the mISs obtained with this thermal Monte
Carlo as a function of the degree $z, K$ in both random regular graphs and ER random graphs.
It is interesting to see that maximal-independent sets found with
standard simulated annealing have different statistical properties
compared to the solutions of greedy algorithms.
Checking on the entropy curves obtained with BP equations, we see that
MC simulations find the thermodynamically
relevant solutions that, in the absence of chemical potential ($\mu = 0$)
are those of typical density that corresponds to the  entropy maximum.
At a difference with Monte Carlo, the Gazmuri's algorithm does not find solutions of typical density but systematically overestimates
it, finding solutions of slightly larger density of coverings. This phenomenon could be due to the non-equilibrium nature of the process and deserves further investigation.

In order to find solutions in the region of non-typical densities, we consider two main strategies:
$i)$ a MC algorithm working at fixed number of covered nodes (fMC); $ii)$ a MC algorithm fixing the density by means of a chemical potential (gMC).

In the first case, it is possible to use the following non-local
Kawasaki-like move: 1) pick up two nodes at random, if they are not
both $0$s or $1$s, exchange them and compute the variation of energy
(number of violated constraints). 2) Accept the move with usual
Metropolis criterion depending on the variation of the energy $\Delta E$
and on the inverse temperature $\beta$. Cooling the system from high
temperature to zero allows to find mIS at a given density of $1$s. In
our simulations performed on graphs of $N=10^3$ nodes, we are able to
find mISs at all densities between a lower limit $\rho_{fMC}^{lower}$
and an upper one $\rho_{fMC}^{upper}$ (see Table \ref{table3}). In Fig. \ref{mc4}, these values
are reported (open triangles) on the RS entropy curve for ER  (left)
and random regular graphs (right). Note that they are quite far from the minimum and maximum predicted by cavity methods.

\begin{figure}
\includegraphics*[width=12cm, angle=0]{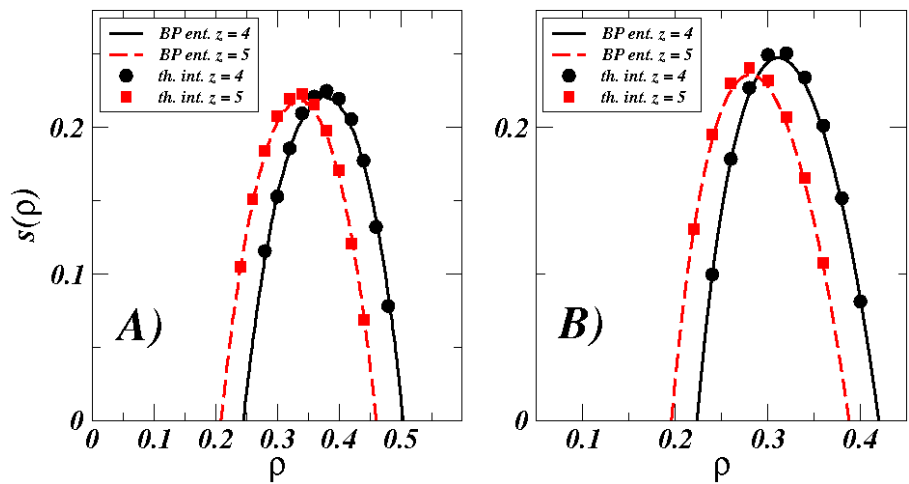}
\caption{(Color online) Entropy curves $s(\rho)$ obtained with BP on both ER (left)
  and random regular graphs (right). On the curves we show the results of the
  thermodynamic integration by Monte Carlo methods (squares and
  circles) averaged over at least $50$ realizations of graphs with
  $N=10^3$ nodes.
} \label{mc2}
\end{figure}

An alternative approach consists in using a grand-canonical lattice
gas formulation, or in terms of spins by the addition of an external
chemical potential coupled with the density $\rho$, i.e. changing the energy $E \to E +\mu \sum_i \sigma_i$.
 The global optimization of the energy now mixes the attempt to
 minimize the number of violated constraints with that of  minimizing
 (or maximizing depending on the sign of $\mu$) the number of covered
 nodes and requires a careful fine tuning of parameters in order to
 get zero violated constraints at the expected density of covered
 nodes.
A better choice is that of modifying the energy as $E \to E +\mu |\sum_i \sigma_i - N \rho^{*}|$, with $\rho^{*}$ being the desired density of covered nodes.  Apart from the details of implementation, the Monte Carlo dynamics follows the usual thermal criterion: 1) pick up a node randomly and flip its binary variable; 2) if the energy is decreasing, then accept the move with probability $1$, otherwise accept the move with a probability $e^{-\beta \Delta E}$.
By fixing $\mu>0$ we just tune the speed of the convergence of the density of $1$s to the desired value $\rho=\rho^*$ during the cooling process (increasing values of $\beta$).

The fact that the number of $1$s is fixed only on average does not seem to help the system
to accomodate the configurations more easily than in the  $fMC$ case. The results are comparable and, in the low density region, $fMC$
seems to perform better (see Fig. \ref{mc4} and Table \ref{table3}).

\subsubsection{Entropy by Thermodynamic Integration}

Monte Carlo algorithms can be used also to give an estimate of the entropy of solutions
by means of the well-known {\em thermodynamic integration method} \cite{LB00}. This method, that is commonly used to
 compute the number of metastable states or blocked configurations in granular systems \cite{BKLS00},  can be applied to the present problem in a very natural way.

For a  system in the canonical ensemble, we can express the specific heat $C$ as a function of the internal  energy $E$, by  $C(\beta) = -\beta^2 \frac{\partial E}{\partial \beta}$, and as a function of the entropy $S$, by $C(\beta) = - \beta \frac{\partial S}{\partial \beta}$. Energy and entropy are therefore related by $d S(\beta) = \beta \frac{\partial E}{\partial \beta} d\beta$.
Since the energy can be computed numerically using a Monte Carlo algorithm, it is convenient to integrate by parts and consider
\begin{equation}\label{MCint}
s(\beta) - s(\beta = 0) =\beta e(\beta) - \int_{0}^{\beta}
e(\beta') d\beta' ,
\end{equation}
where we have used the rescaled quantities $s=S/N$ ans $e=E/N$.
Equation \ref{MCint}  provides the zero-temperature entropy $s(\infty)$ once we know $e(\beta)$ and the infinite temperature entropy $s(0)$.
In our case, the calculation gives an estimate of $e(\beta)$ and so the entropy of maximal-independent sets on a given graph.
Indeed using Monte Carlo we can not find very accurate values for the average energy expecially if there exist a phase transition.
Even when there is no phase transition taking place in the integration
range, there are other sources of inaccuracy. More precisely,  we can
investigate a large but finite interval $[0,\beta_{max}]$, with
$\beta_{max} \ll \infty$, thus if the Monte Carlo algorithm is not
able to reach the ground-states at some $\beta \in [0,\beta_{max}]$,
the numerical integration can only provide a upper bound for the real entropy.
In some situations, the two errors may sum up, because replica
symmetry breaking also causes slowing down in Monte Carlo
algorithms. This is the main reason why we cannot push this method up
to the extreme values of density predicted by theoretical
calculations.

We have used the grand-canonical MC algorithm (gMC) described before to
compute the entropy of solutions with non typical densities of covered
nodes. Note that the integration formula Eq. \ref{MCint} is correct
also for the gMC algorithm because it corresponds to a canonical spin
system with external field $h \propto \mu/\beta$, that only modifies
the energetic contribution. At infinite temperature, $\beta=0$,  the
entropy is $\log(2)$ because all states are accessible, but for large
values of the chemical potential $\mu$, the system rapidly
concentrates around the desired density $\rho$ as soon as we increase
$\beta$. The final entropy $s(\beta_{max})$ gives an estimate of the
entropy of mIS with a density $\rho$ of covered nodes. In
Fig. \ref{mc2} we show the results of thermodynamic integration for
both ER and random regular graph with $z,K=4$ and compare them with
the curves obtained with the cavity approach. We have averaged the
entropy values over $50$ realizations of the graphs and the standard deviation
of the values are smaller than data symbols. The points perfectly
agree with the results of the cavity approach showing that, in the
range of validity of BP equations they correctly predict the
statistical properties of maximal-independent sets.

\begin{figure}
\includegraphics[width=10cm,angle=0]{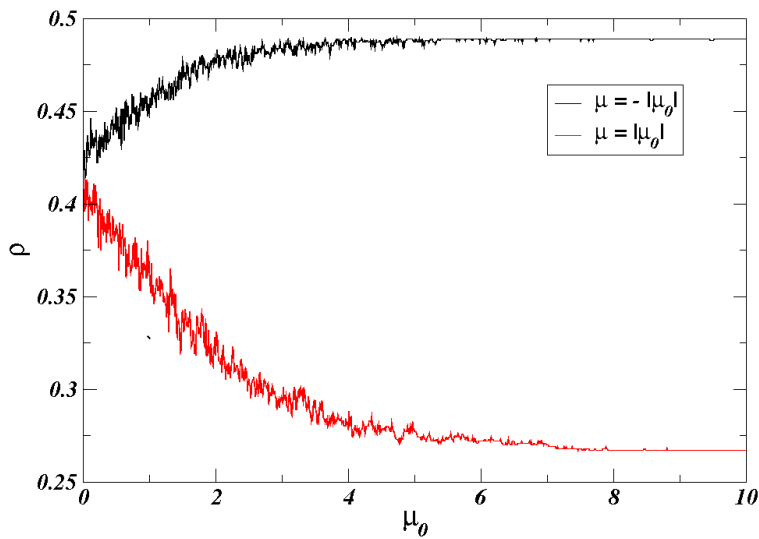}
\caption{(Color online) Example of the slowing down phenomenon taking place at large
  chemical potential in the ``rearrangement
  Monte Carlo'' method explained in Section \ref{walkSEC}. The curves
  represent the density of covered nodes in the mIS sampled by the rMC
  algorithm as chemical potential $\mu$ slowly increases (lower curve) or decreases (upper curve) in an ER graph of
  $N=10^3$ nodes and average degree $z=4$.
} \label{mc3}
\end{figure}

\subsubsection{Walking on the space of solutions by ``rearrangement
  Monte Carlo''}\label{walkSEC}

In Section \ref{orgSEC} we have seen that it is possible to go from a
mIS to another one repeating a simple operation, that consists in
flipping a variable from $0$ to $1$ (or viceversa) and rearranging the
values of all neighbors iteratively until a new mIS is found (see
Fig. \ref{diag2}).  It was proved that the operation always involves a finite number of variables, that makes it possible to implement this process inside a Monte Carlo algorithm.
Moreover, Proposition \ref{prop4} ensures that, given two mIS
configurations, it is always possible to go from one to the other and
back with a sequence of operations of this kind. The sequence of
operations is finite whenever $N$ is finite.
A Monte Carlo algorithm based on this operation is thus expected to be
ergodic in the space of all maximal independent sets (see App. \ref{appORG} and Ref. \cite{DPR09}).

We define the following MC algorithm:
\begin{itemize}
\item[0.] The initial state is chosen finding a typical mIS by best-response starting from a random configuration.
\item[1.] We select a node randomly, we try to flip it and readjust all neighboring nodes propagating the rearrangement until all nodes are satisfied. In this way we generate another mIS.
\item[2.] We compute the variation $\Delta \rho$ of the density of covered nodes between the two configurations, and accept the move with probability $1$ if the density decreases and probability $e^{-\mu \Delta \rho}$ if it increases.
\item[3.] We repeat points 1.-3. for a given number of iterations,
  then we stop or change the chemical potential $\mu$.
\end{itemize}
By performing a simulated annealing in which $\mu$ is slowly increased
(decreased) from $0$, we find a chain of maximal independent sets with
decreasing (increasing) density of covered nodes.

Having finite rearrangements means that the variations in the number of covered nodes are finite as well and the density $\rho$ is almost constant. Therefore appreciable density fluctuations only occur after $O(N)$ rearrangements, that is a Monte Carlo step.
When $\mu$ is varied at a sufficiently slow rate, the algorithm should be able to find mISs at all densities at which they exist.
Fig. \ref{mc3} shows some data points taken every $100$ MC
steps. The density of the mISs sampled by the algorithm is reported as
function of $\mu$. The curves seem to converge
to values of the density that are still far from the two theoretical
bounds (obtained by the cavity method) of the lower and upper
SAT/UNSAT transitions. At low and large density values, the
algorithm is not able to find mIS beyond some threshold
$\rho_{rMC}^{lower}>\rho_{BP}^{lower}>\rho_{min}^{lower}$ and $\rho_{rMC}^{upper}<\rho_{BP}^{upper}<\rho_{max}^{upper}$.
Fig. \ref{mc4} reports these two values (blue circles) in ER and
random regular graphs for average degree $4$. A direct comparison with other
computational bounds shows that this algorithm outperforms the other Monte
Carlo methods in finding mISs at low and high density of covered nodes.
In particular it is much faster than the other MC algorithms and provides a large number of mIS at different
densities in a reasonably short time.
The obtained bounds are quite worse than those obtained by BP
decimation in the low density phase, but they are better in the high-density phase (see Table \ref{table3} and Fig. \ref{mc4}).

The exact value of the density at which the algorithm stops is only
estimated by several numerical experiments and further investigation
is required in order to understand if some heuristic optimization
could allow to reach better results, even in the presence of replica symmetry breaking.
In fact, at very large chemical
potential the algorithm becomes sensitive to very small barriers, due
to the flip of a finite number of variables. Such barriers do not
require a change in density, but can trap the algorithm in local minima if the algorithm is running at very large chemical potential.
If this is true, some heuristic method could be designed in
order to improve the performances of the rMC algorithm.

\subsection{BP decimation}\label{BPdecSEC}

Given an instance of random graph we can run BP equations in Eq.
\ref{BPeq1} starting from random initial values for messages
$\underline{r}^{i \to j}$. If we reach a fixed point of the
equations then the local marginals

\begin{eqnarray}
b_i= \frac{e^{-\mu}\prod_{j\in \partial i}(1-r_1^{j \to i})}{e^{-\mu}\prod_{j\in \partial i}(1-r_1^{j \to i})+
\prod_{j\in \partial i}(1-r_{00}^{j \to i})-\prod_{j\in \partial i} r_0^{j \to i}},
\end{eqnarray}

will give us the approximate probability of $\sigma_i=1$
among the set of mIS's. One strategy of finding a mIS
is to decimate the most biased variables according to their
preference. Suppose that in the first run of algorithm variable
$i$ has the maximum bias $|1-2b_i|$ among the variables.
Then if, for example, $b_i>1/2$ we fix $\sigma_{i}=1$ and
reduce the problem to a simpler one with
smaller number of variables. The strategy in BP decimation
algorithm \cite{BMZ05} is to iterate the above procedure till we
find a configuration of variables that satisfies all the
constraints. Certainly if the believes $b_i$ that we
obtain are exact the algorithm would end up with a mIS, if there
exist any. Otherwise at some point we would find contradictions
signaling the wrong decimation of variables in previous steps.

The results of BP decimation algorithm have been summarized in
Table \ref{table3}. For the case $K,z = 4$, the minimum and
maximum density at which we are able to find a mIS for graphs of
size $N=10^4$ are also reported as black full circles in Figure
\ref{mc4}.

Notice that similarly one can use a SP decimation algorithm based on
SP equations, to find a solution (here a mIS). This is usually more useful
than BP decimation in problems that exhibit a well clustered solution space.
In the case of maximal independent sets we did not observe a significant difference
in the performance of the two algorithms. This is the reason why here we  focus
on the BP decimation algorithm which is more accessible.

\begin{table}
\begin{center}

\begin{tabular}{|c||c|c|c|c|c||c|c|c|c|c|}
  \hline

    $K$      &  $\rho_{min}^{BP}$ & $\rho_{min}^{BPD}$ &
    $\rho_{min}^{rMC}$  &  $\rho_{min}^{gMC}$ & $\rho_{min}^{fMC}$ & $\rho_{max}^{fMC}$ &  $\rho_{max}^{gMC}$  & $\rho_{max}^{rMC}$ & $\rho_{max}^{BPD}$ & $\rho_{max}^{BP}$     \\
  \hline

    $3$      & $0.264$ & $0.267$   & $0.269$  & $0.275$ &  $0.271$ & $0.449$ & $0.449$ & $0.449$  & $0.449$ & $0.458$    \\

  \hline

    $4$      & $0.223$ & $0.228$   & $0.230$ &  $0.235$ & $0.232$ &  $0.408$  & $0.408$
    & $0.410$ & $0.408$  & $0.419$  \\

  \hline

    $5$      & $0.196$ & $0.201$  & $0.203$  & $0.206$ & $0.206$ & $0.377$ & $0.377$  & $0.379$ & $0.375$   & $0.387$  \\

  \hline

    $6$      & $0.175$  & $0.181$  & $0.185$ & $0.188$ & $0.184$ & $0.349$ & $0.349$ & $0.349$ & $0.348$   & $0.360$  \\

  \hline

    $7$      & $0.159$ & $0.165$   & $0.168$ & $0.172$ & $0.169$ & $0.327$ & $0.327$ & $0.328$ & $0.325$   & $0.338$  \\

  \hline

    $8$      & $0.146$ & $0.153$   & $0.156$ & $0.158$   & $0.158$ & $0.308$ & $0.310$  & $0.308$  & $0.306$   & $0.319$  \\

  \hline

    $9$      & $0.136$ & $0.143$   & $0.145$ & $0.148$  & $0.146$ & $0.294$ & $0.293$ & $0.294$  & $0.289$   & $0.301$  \\

  \hline

    $10$      & $0.127$ & $0.134$   & $0.138$  & $0.140$  & $0.138$ & $0.278$ & $0.278$ & $0.278$ & $0.274$  & $0.287$ \\

  \hline

\end{tabular}

\vskip 0.5cm

\caption{Summary of the minimum and maximum values of density at which
  we find mISs on random regular graphs with different algorithms: BP
  decimation $\rho_{min}^{BPD}$, $\rho_{max}^{BPD}$; fixed-density
  Monte Carlo $\rho_{min}^{fMC}$, $\rho_{max}^{fMC}$; grand-canonical
  Monte Carlo $\rho_{min}^{gMC}$, $\rho_{min}^{gMC}$; rearrangement
  Monte Carlo $\rho_{min}^{rMC}$, $\rho_{min}^{rMC}$. BP results are
  obtained on graphs of size $N=10^4$ whereas Monte Carlo results on
  graphs of size $N=10^3$. In the cases of $fMC$ and $gMC$, we have chosen the values at which at least half of the runs were successful in
  finding a mIS.}\label{table3}
\end{center}
\end{table}

\section{Conclusions and Outlook}

In this paper we have investigated the statistical properties of
maximal independent sets, a graph theoretic quantity that plays a
central role both in combinatorial optimization and in game
theory. Among the most prominent applications it is worth
mentioning the development of distributed algorithms for radio
networks \cite{MW05} and the study of public goods allocation in
economics \cite{GGJVY08}.

A long-standing problem in combinatorial optimization is to
estimate the number of maximal independent sets in a given graph,
and devise efficient algorithms to find them, independently of
their size. In the first part of the paper we have focused on some
theoretical methods to compute the number of mISs of size $M$ in
random graphs of size $N$. As in general this number is
exponentially large $\mathcal{N}_{mIS} \approx e^{N s(M/N)}$, we
have used statistical mechanics methods to compute the entropy
$s(\rho)$ of maximal independent sets as a function of the density
$\rho = M/N$ of coverings and of the average degree  of the
graphs. At typical density values, the RS approximation (BP
equations) describes correctly the system. While the BP equations remain
stable in the low density region, for high density
of coverings the BP equations become unstable
and the RS solution does not hold anymore.

The general 1RSB calculations for
random regular graphs show a dynamical transition in a small interval
of density very close to the minimum density. However, the population dynamics
algorithm hardly converges to this solution.
In the high density region we observe a condensation
transition to 1RSB phase.
Here the solution has better convergence than the 1RSB solution for low densities.
Previous studies, for instance in $3$-SAT problem \cite{MRS08}, show
that these solutions suffer from another kind of instability.
A more detailed study of stability of 1RSB solutions in this problem
remains to check for future works.

From the computational point of view, the main issue is to find
maximal independent sets at very low or very high density, within
the bounds indicated by RS and  first-moment calculations.
Greedy algorithms, like Gazmuri's one, can find mISs at very
typical values of the density, whereas Monte Carlo methods and BP
decimation can be used to explore regions of non-typical values.
Our numerical calculations indicate that BP decimation gives the
best performances, but the values at which we find solutions are
still far from the theoretically predicted bounds. Notice that
despite the dynamical transition in the low density region, the absence of
globally frozen variables could make the problem easy on average in that region
\cite{S08,DRZ08}.

We expect that the results we obtained here could be exploited to
design more efficient algorithms. A first example is represented
by the {\em rearrangement Monte Carlo} algorithm, that allows to
move among the space of mISs  sampling them with a
density-dependent Gibbs measure. Apart from dramatic slowing down
faced by MC algorithms in presence of RSB, the algorithm should be
able in principle to reach all existing mISs.

A maximal independent set on a graph $\mathcal{G}$ can be viewed as
a saturated packing of hard spheres of diameter $d=2$. So the minimum
and maximum mIS densities define the region one can have saturated packings
and in this paper we gave these limits in the RS approximation. It would be interesting
to see how computational and physical properties of mISs change by increasing diameter $d$.

\acknowledgments

We would like to thank M. Marsili, F. Zamponi, L. Zdeborov\'a and R. Zecchina for useful discussions and comments.
P. P. acknowledges support from the project Prin 2007TKLTSR
"Computational markets design and agent--based models of trading behavior".

\appendix

\section{Annealed calculations and bounds in random graphs}
\label{appANN}

 We  compute here some rigorous mathematical results on the number of maximal independent sets with a given density $\rho$ of covered nodes. The first moment method allows to give lower bounds $\rho_{min}^{lower}$ and upper bounds $\rho_{max}^{upper}$ for the density of covered nodes in a mIS on random graph.

Let $X_{M}$ denote the number of mISs of size $M$ in a graph $\mathcal{G}$ of size $N$, from the Markov inequality we have
\begin{eqnarray}
Prob(X_{M}>0) \le \overline{X}_M.
\end{eqnarray}
where $\overline{X}_M$ is the average number of mISs in the ensemble of graphs of size $N$ to which $\mathcal{G}$ belongs.
If for some values of $M < N$ the average number of maximal independent sets of size $M$ becomes zero, then Markov inequality implies that the probability to find a mIS of size $M$ also vanishes.

In the Erd\"os-R\'enyi ensemble of random graphs $G(N,p)$, the average number of maximal independent sets of size $M$ is given by \cite{BF04}
\begin{eqnarray}
\overline{X}_M=\binom{N}{M}(1-p)^{\frac{M(M-1)}{2}}[1-(1-p)^M]^{N-M}.
\end{eqnarray}
For large $N$ and $M$ with fixed $\rho=M/N$ and $p=z/N$,  the asymptotic behavior of the number of mIS is $\overline{X}_M \approx e^{N s_1(\rho)}$ where
\begin{eqnarray}
s_1(\rho)=-\rho \ln(\rho)-(1-\rho) \ln(1-\rho)-\frac{1}{2}z
\rho^2+(1-\rho)\ln(1-e^{-z\rho}).
\end{eqnarray}
The density values  at which the entropy $s_1$ becomes negative give bounds for the existence of maximal independent sets. Therefore
$\rho_{min}^{lower}$ is a lower bound for the density of occupied nodes in the minimum mIS (mis) and $\rho_{max}^{upper}$ is an upper bound for the density of occupied nodes in the maximum mIS (MIS).
The values of $\rho_{min}^{lower}$ and $\rho_{max}^{upper}$ for ER random graphs with several values of the average degree $z$ are reported in the first and last column of Table \ref{table1}.

The first moment calculation can be extended to random regular graphs (RRG), i.e. graphs in which connections are established in a completely random way with the only constraint that all nodes have the same finite degree $K$ (we will consider diluted networks, i.e. $K \ll N$)

It gives
\begin{eqnarray}
\overline{X}(\rho) & \simeq &
\binom{N}{\rho N}\left[ 1 - \rho^2 \right]^{\frac{K N}{2}} \left[ 1 -  (1- \rho)^{K} \right]^{(1-\rho) N} \\
& \approx  & \exp{N  \left[ -\rho \ln(\rho)-(1-\rho) \ln(1-\rho) - \frac{K}{2} \rho^2+ (1-\rho) \log{\left[ 1 -  (1- \rho)^{K} \right]}\right]} .
\end{eqnarray}
As before, extracting the zeros of the entropy function for various
degrees $K$, we obtain the values of $\rho_{min}^{lower}$ and
$\rho_{max}^{upper}$ that are reported in Table \ref{table2}. As we will
verify later comparing these results with those from the cavity method, only the lower bound is tight, whereas the upper one
strongly overestimates the existence of mISs.  However, the entropy
values at typical density of coverings (e.g. the maximum of the entropy) are in agreement with other theoretical and numerical results, corroborating the validity of this improved annealed calculation.

The annealed approximation we have just discussed only gives an upper bound for the real number of mISs, therefore it would be important to have also a lower bound for the real entropy curve $s(\rho)$. This is usually obtained by the second moment method.
Let $\overline{X_M^2}$ be the second moment of the number of mISs of size $M$ in a graph of size $N$,
the Chebyshev's inequality provides a lower bound for the probability of finding a mIS of size $M$,
\begin{eqnarray}
Prob(X_{M}>0) \ge  \frac{\overline{X}_M^2}{\overline{X_M^2}}.
\end{eqnarray}
For ER random graphs, the second moment is
\begin{eqnarray}
\overline{X_M^2}=\sum_{l=0}^M \binom{N}{M}\binom{M}{l}\binom{N-M}{M-l}
(1-p)^{M(M-1)-\frac{l(l-1)}{2}}[1-(1-p)^{2M-l}]^{N-(2M-l)}
\end{eqnarray}
where $l$ is the overlap between the two configurations corresponding to mISs of size $M$. In the
scaling limit, $\rho=M/N$ and $x = l/N$
\begin{eqnarray}
\overline{X_M^2}=N\int_{0}^{\rho} dx e^{Ns_2(\rho, x)}.
\end{eqnarray}
Let us call $x^*$ the value of the overlap maximizing $s_2(\rho,x)$.
The density values where $2 s_1(\rho)-s_2(\rho,x^*)$ vanishes should give the lower bound $\rho_{max}^{lower}$
for the density of the MIS and the upper bound $\rho_{min}^{upper}$ for the density of the mis.
Unfortunately, for ER random graphs, $s_2(\rho,x^*)$ is always larger than $2 s_1(\rho)$, meaning that the ratio
$\overline{X}_M^2 / \overline{X_M^2}$ always vanishes in the thermodynamic limit. The corresponding
 trivial result $Prob(X_M >0) \geq 0$ does not say anything about the extremal densities for MIS and mis.
The same holds for the ensemble of random regular graphs.

\section{Proof of the results of Section \ref{orgSEC}}
\label{appORG}

Here are the proofs of the result shown in Section \ref{orgSEC}, from which we maintain the notation.
Consider a finite network and call $\sigma_i \in \{ 0,1 \}$ the membership of node $i$ to a set $\mathcal{I}$.
It is clear that there is a one--to--one correspondence between any subset of the nodes and any vector $\underline{\sigma}$.
We will consider those $\underline{\sigma}$ for which $\mathcal{I}$ is a mIS in a given graph $\mathcal{G}$.
Call finally $N^1_i$ the set of nodes which are first neighbors of node $i$, and $N^2_i$ those which are second neighbors of node $i$.
By definition of mIS, we have that  $\mathcal{I}$ is a mIS if and only if $\underline{\sigma}$ is such that
\begin{equation}
\left\{
\begin{array}{ll}
\sigma_i=1 & \mbox{if $\sigma_j=0$ for any neighbor $j$ of node $i$;}  \\
\sigma_i=0 & \mbox{otherwise.}
\end{array}
\right.
\label{rule}
\end{equation}
Equation \ref{rule} defines what is the best response dynamics.
For any node $i$ there is always one strict best response given any memberships' configuration of its neighbors in $N_i^1$. This would hold 'a fortiori' also in a mIS configuration, and hence proves that $\underline{\sigma}$ is locally frozen (Proposition \ref{prop1}).

\bigskip

Proposition \ref{prop3} tells us that the best response rule will imply a new mIS, and that any best response dynamics of the other nodes will be limited to the second degree neighborhood of the node which initially flipped.

\bigskip

{\bf Proof of Proposition \ref{prop3}:}
suppose node $i$ is in the mIS, so that $\sigma_i=1$, and we remove it so that $\sigma_i^{new}=0$.
Consider now any node $j$ in $N^1_i$, it is clear that $\sigma_j=0$ since $\sigma_i=1$.
By best response, for all those  $j \in N^1_i$ such that $\sigma_k=0$ for any $k \in N^1_j \backslash \{ i \}$, we will have
$\sigma_j^{new}=1$.
In the case that two such $j$'s that flipped from $0$ to $1$ will be linked together, by best response only some of them will flip to $1$ (this is the only random part in the best response rule).
If $j$ is such that $\sigma_j=0$ and $\sigma_j^{new}=1$, it is surely the case that any $k \in N^1_j \backslash \{ i \}$ was playing $\sigma_k=0$ and remains at $\sigma_k^{new}=0$.
Finally, if no neighbors $j \in N_i$ flip from $0$ to $1$, we will allow node $i$ to turn back to its original position.
The propagation of the best response is then limited to $N^1_i \cup \{ i \}$ (and ends in an mIS, possibly the old one).

\smallskip

{\bf Note:} \emph{a best response from $0$ to $1$ applies only to nodes that are $0$, are linked to a node which is flipping from $1$ to $0$, and that node is the only neighbor they have who is originally $1$.}

\smallskip

Suppose now that $\sigma_i=0$ and we flip it so that $\sigma_i^{new}=1$.
The nodes $j$ in $N^1_i$ who had $\sigma_j=0$ will continue to do so.
Any node $j$ in $N^1_i$ (at least one) who had $\sigma_j=1$ will move to $\sigma^{new}_j=0$.
By the previous point this will create a propagation to some $k \in N^1_j$, but not $i$.
This proves that the propagation of the best response is limited to $N^1_i \cup N^2_i$ (and ends in a new mIS). \fine

\bigskip

Finally, we prove that any mIS can be reached in finite steps, by best response dynamics, from any other mIS (Proposition \ref{prop4}).

\bigskip

{\bf Proof of Proposition \ref{prop4}:}
we proceed by defining intermediate mIS $\underline{\sigma}^1$, $\underline{\sigma}^2, \dots$
between any two mIS $\underline{\sigma}$ and $\underline{\sigma}'$ (associated to $\mathcal{I}$ and $\mathcal{I}'$).
$\underline{\sigma}^{n+1}$ will be obtained from $\underline{\sigma}^n$ by flipping one node from $0$ to $1$ and waiting for the best response of all the others.

\smallskip

If two mIS $\underline{\sigma}$ and $\underline{\sigma}'$ are different, it must be that there is at least one $i_1$ such that $\sigma_{i_1}=0$ and $\sigma'_{i_1}=1$ (by definitions any strict subset of a maximal independent set is not a covering any more).
Change the membership of that node so that $\sigma_{i_1}^{1}=\sigma'_{i_1}=1$.
By previous proof this will propagate deterministically to $N^1_{i_1}$ and, for all $j \in N^1_{i_1}$, we will have $\sigma^{1}_j = \sigma'_j = 0$.
Propagation may also affect $N^2_{i_1}$ but this is of no importance for our purposes.

\smallskip

If still $\underline{\sigma}^{1} \ne \underline{\sigma}'$, then take another node $i_2$ such that $\sigma^1_{i_2}=0$ and $\sigma'_{i_2}=1$ (${i_2}$ is clearly not a member of $N^1_{i_1} \cup \{ i_1 \}$).
Pose $\sigma_{i_2}^{2}=\sigma'_{i_2}=1$, this will change some other nodes by best response, but not $j \in N^1_{i_1} \cup \{ i_1 \}$, because any
$j \in N^1_{i_1}$ can rely on $\sigma^{1}_{i_1}=1$, and then also $\sigma^{2}_{i_1}=\sigma^{1}_{i_1}=1$ is fixed.

\smallskip

We can go on as long as $\underline{\sigma}^{n} \ne \underline{\sigma}'$, taking any node $i_{n+1}$ for which $\sigma^n_{i_{n+1}}=0$ and $\sigma'_{i_{n+1}}=1$.
This process will converge to $\underline{\sigma}^{n} \rightarrow \underline{\sigma}'$ in a finite number of steps because:
\begin{itemize}
\item when $i_{n+1}$ shifts from $0$ to $1$, the nodes $j \in \bigcup_{h=1}^n \left( N^1_{i_h} \cup \{ i_h \} \right)$ will not change, since they are either $0$--nodes with a $1$--node beside already (the $1$--node is some $i_h$, with $h \leq n$), or a $1$ (some $i_h$) surrounded by \emph{frozen} $0$'s;
\item by construction it is never the case that $i_{n+1} \in \bigcup_{h=1}^n \left( N^1_{i_h} \cup \{ i_h \} \right)$, because for all $j \in \bigcup_{h=1}^n \left( N^1_{i_h} \cup \{ i_h \} \right)$ we have that $\sigma^n_{j} = \sigma'_j$;
\item the network is finite. \fine
\end{itemize}

\bigskip

The shift from $\underline{\sigma}$ to $\underline{\sigma}'$ is done by construction re--defining the covering of any $\underline{\sigma}^n$ from the covering of $\underline{\sigma}'$.
It is always certain that, by best response, any $\underline{\sigma}^n$ is also an independent set.

\section{Survey Propagation}
\label{appSP}

Survey Propagation (SP) \cite{MeZ02} allows to study the 1RSB phase (if any) in a
simplifying limit where we have only one parameter $y = m \mu$.
The corresponding equations are called SP equations and provide
the behavior of the $m=0$ complexity $\Sigma (\rho)$.

If the solutions are clustered, the believes $\{r_{1}^{i \to j},r_{0}^{i
  \to j}, r_{00}^{i \to j}\}$ are not distributed in the same way from
cluster to cluster (different Gibbs states), therefore we have to
introduce a distribution of cavity fields. In writing SP equations we assume
that these distributions are $\mathcal{P}[\underline{r}^{i \to j}] =
 \sum_a \eta_{a} \delta(r_{a}^{i \to j}-1) \prod_{b\ne a}\delta(r_{b}^{i \to j})$ for $a,b = 0,00,1$.
Cavity equations thus translate into equations for the surveys
$\{\eta_{0}, \eta_{00}, \eta_{1} \}$.

In the limit $\mu \to -\infty$ they read
\begin{eqnarray}\label{SP1}
\eta_1 &= & \frac{e^{-y}(1-\eta_1)^{K-1}}{1+(e^{-y}-1)(1-\eta_1)^{K-1}},\\ \nonumber
\eta_0 &= &\frac{ 1-(1-\eta_1)^{K-1}}{1+ (e^{-y}-1)(1-\eta_1)^{K-1}},\\ \nonumber
\eta_{00} &=&0,
\end{eqnarray}
where $e^{-y}$ is the penalty favoring clusters (solutions) with
higher density.
The density of occupied nodes is

\begin{equation}
\rho = \frac{e^{-y}(1-\eta_1)^{K}}{1+(e^{-y}-1)(1-\eta_1)^{K}}.
\end{equation}

In the Bethe approximation
\begin{equation}
y \Phi = y \Delta \Phi_i-\frac{K}{2}y \Delta \Phi_{ij},
\end{equation}

with
\begin{eqnarray}
e^{-y \Delta \Phi_i} & =&1+(e^{-y}-1)(1-\eta_1)^K,\\ \nonumber
e^{-y \Delta \Phi_{ij}} &= &\eta_0^2+2\eta_1\eta_0,
\end{eqnarray}

\begin{figure}
\includegraphics[width=10cm,angle=0]{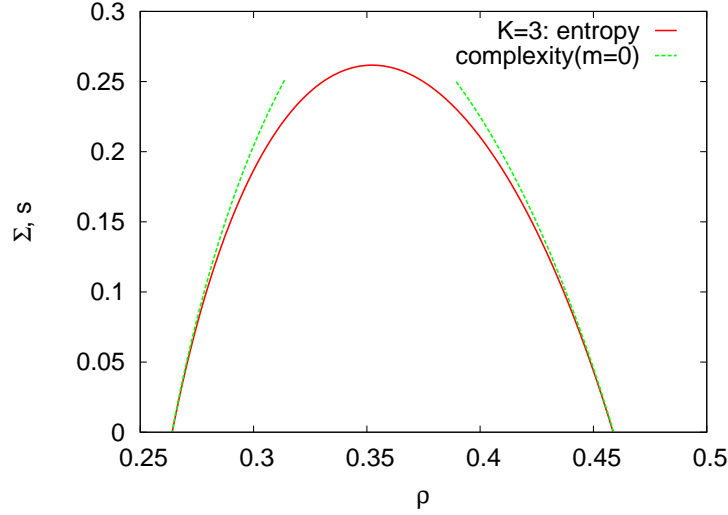}
\caption{(Color online) Comparing $1$RSB complexity for $m=0$ (dashed line) with the
BP entropy (solid line).}\label{complexitym0}
\end{figure}

Solving numerically the equations we find that the complexity is always a bit larger than the
BP entropy, that is an unphysical result. Figure \ref{complexitym0} compares the complexity with the RS entropy for $K=3$.
The same behavior is observed for larger degrees. Indeed, looking at the stability of this
first set of SP equations (details are reported in Appendix
\ref{appSA}), it turns out that the SP equations \ref{SP1} are
not stable in the whole large density region.

In the other limit $\mu \to + \infty$, we find the equations
\begin{eqnarray}\label{SP2}
\eta_1 & \propto & e^{-y} \left[(1-\eta_1)^{K-1} -
  \eta_{0}^{K-1}\right] \\
\eta_0 & \propto & (1-\eta_{00})^{K-1} - \eta_0^{K-1} \\
\eta_{00} & \propto & \eta_0^{K-1}
\end{eqnarray}
while the generalized thermodynamic potential is given by the same
expression but with
\begin{eqnarray}
e^{-y \Delta \Phi_i} & =&e^{-y}(1-\eta_1)^K + (1-\eta_{00})^K - \eta_0^{K},\\ \nonumber
e^{-y \Delta \Phi_{ij}} &= &\eta_0^2+2\eta_1(\eta_0+\eta_{00}),
\end{eqnarray}
with density $\rho$ obtained as

\begin{equation}
\rho = \frac{e^{-y}(1-\eta_1)^{K}}{
e^{-y}(1-\eta_1)^K+(1-\eta_{00})^{K}-\eta_0^K }.
\end{equation}

In Appendix \ref{appSA} we show that the SP solution is stable in the whole low-density
region. However as in the high density region we find a complexity that is a bit larger than the RS entropy, see figure \ref{complexitym0}.

As expected, $m=0$ is not the correct approximation to describe our system in the 1RSB phase.

\section{Stability Analysis} \label{appSA}
\subsection{RS stability}

The BP equations for the mIS problem involve three cavity fields $r_a^{i
  \to j}$ with $a=1,0,00$, therefore in order to check the
stability of the RS solution we have to compute the response
induced in these fields by a small perturbation in the neighboring
cavity fields $r_a^{k\to i}$. The elements of the stability matrix
are obtained as  $M_{a,b} = \frac{\partial r_{a}^{i \to
j}}{\partial r_{b}^{k \to i}}$,
\begin{eqnarray}\label{stabmatrix}
M_{1,1} & = &
-\frac{e^{-\mu}(1-r_1)^{K-2}}{Z_M}+(1-r_1)\left[\frac{e^{-\mu}(1-r_1)^{K-2}}{Z_M}\right]^2,\\
\nonumber
M_{1,0} & = & 0,\\ \nonumber
M_{1,00} & = & (1-r_{00})^{K-2}\frac{e^{-\mu}(1-r_1)^{K-1}}{Z_M^2},\\
M_{0,1} & = &
\left[(1-r_{00})^{K-1}-r_0^{K-1}\right]\frac{e^{-\mu}(1-r_1)^{K-2}}{Z_M^2},\\
\nonumber
M_{0,0} & = & -\frac{r_0^{K-2}}{Z_M},\\ \nonumber
M_{0,00} & = & -\frac{(1-r_{00})^{K-2}}{Z_M}+(1-r_{00})^{K-2}\frac{(1-r_{00})^{K-1}-r_0^{K-1}}{Z_M^2},\\
M_{00,1} & = & r_0^{K-1}\frac{e^{-\mu}(1-r_1)^{K-2}}{Z_M^2},\\
\nonumber M_{00,0} & = & \frac{r_0^{K-2}}{Z_M},\\ \nonumber
M_{00,00} & = & (1-r_{00})^{K-2}\frac{r_0^{K-1}}{Z_M^2},
\end{eqnarray}
with $Z_M=e^{-\mu}(1-r_1)^{K-1}+(1-r_{00})^{K-1}$.

If $\lambda_{M}$ is the largest eigenvalue of $\mathbf{M}$,
computed at the fixed point of the BP equations, then the
(spin-glass) stability condition reads $\Lambda \equiv
(K-1)\lambda_M^2<1$. If $\Lambda$ is larger than $1$, then the
perturbation is amplified by iteration and the RS solutions are
unstable fixed points of the BP equations.
Figure \ref{stability-SP} shows how $\Lambda$ changes with density in random regular graphs of degree $K=3$.
The same behavior is observed for other degrees.

\begin{figure}
\includegraphics[width=10cm,angle=0]{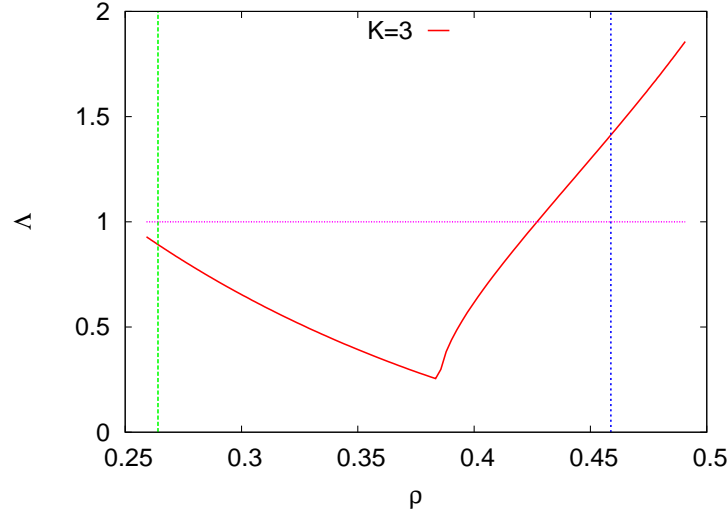}
\caption{(Color online) Checking stability of BP equations: typical behaviour of $\Lambda$ (solid line). Vertical dashed lines represent the minimum and maximum densities predicted by BP equations.}\label{stability-BP}
\end{figure}

\subsection{SP stability}

Suppose that, according to 1RSB picture, solutions are organized in a large number
of clusters that represent different Gibbs pure states. There are two kinds of possible
instabilities: a) states can aggregate into different clusters, or b)
each state can fragment in different states. We study here if this can
occur for the mIS problem in the two limits defining the SP approximation.

\subsubsection{Limit $\mu \to  + \infty$, $m \to 0$ ($y = m\mu$)}

The first kind of instability is related to the divergence of the
inter-cluster spin-glass susceptibility, i.e. the instability of
SP fixed points on single graphs. Hence the calculation is
equivalent to the RS case. We need the $3\times 3$ matrix
$M_{a,b}=\frac{\partial \eta^{i\rightarrow j}_a}{\partial
\eta^{k\rightarrow i}_b}$, where $a=1,0,00$. We have
\begin{eqnarray}
M_{1,1} & = &
-\frac{e^{-y}(1-\eta_1)^{K-2}}{Z_M}+\frac{e^{-y}((1-\eta_1)^{K-1}-\eta_0^{K-1})}{Z_M^2}[e^{-y}(1-\eta_1)^{K-2}],\\
\nonumber
M_{1,0} & = &
-\frac{e^{-y}\eta_0^{K-2}}{Z_M}+\frac{e^{-y}((1-\eta_1)^{K-1}-\eta_0^{K-1})}{Z_M^2}[e^{-y}\eta_0^{K-2}],\\
\nonumber
M_{1,00} & = &
\frac{e^{-y}((1-\eta_1)^{K-1}-\eta_0^{K-1})}{Z_M^2}[(1-\eta_{00})^{K-2}], \\
M_{0,1} & = &
\frac{(1-\eta_{00})^{K-1}-\eta_0^{K-1}}{Z_M^2}[e^{-y}(1-\eta_1)^{K-2}],\\
\nonumber
M_{0,0} & = &
-\frac{\eta_0^{K-2}}{Z_M}+\frac{(1-\eta_{00})^{K-1}-\eta_0^{K-1}}{Z_M^2}[e^{-y}\eta_0^{K-2}],\\
\nonumber
M_{0,00} & = & -\frac{(1-\eta_{00})^{K-2}}{Z_M}+\frac{(1-\eta_{00})^{K-1}-\eta_0^{K-1}}{Z_M^2}[(1-\eta_{00})^{K-2}],\\
\nonumber
M_{00,1}& = &
\frac{\eta_0^{K-1}}{Z_M^2}[e^{-y}(1-\eta_1)^{K-2}],\\ \nonumber
M_{00,0} & =&
\frac{\eta_0^{K-2}}{Z_M}+\frac{\eta_0^{K-1}}{Z_M^2}[e^{-y}\eta_0^{K-2}],\\
\nonumber
M_{00,00}& =&
\frac{\eta_0^{K-1}}{Z_M^2}[(1-\eta_{00})^{K-2}],
\end{eqnarray}
with
\begin{eqnarray}
Z_M=e^{-y}[(1-\eta_1)^{K-1}-\eta_0^{K-1}]+(1-\eta_{00})^{K-1}.
\end{eqnarray}

If $\lambda_M$ is the dominant eigenvalue of $\mathbf{M}$ then the first kind stability condition reads

\begin{eqnarray}
\Lambda_1= (K-1) \lambda_M^2 <1.
\end{eqnarray}

The second kind instability is instead related to intra-cluster
susceptibility, and can be studied by means of ``bug proliferation''
defined on clusters. This means that we consider how a change in a
single cavity field from, say, $a$ to $b$ is propagated to neighbors
and how this reflects on the distribution of surveys inside the
cluster. If the instability ``bug'' is
$\lambda_{a \to b}$, we need the transfer matrix $T_{ab,cd}$ satisfying to
$\lambda_{a\rightarrow b}=\sum_{cd} T_{ab,cd} \lambda_{c\rightarrow
  d}$. The iterative equations are
\begin{eqnarray}
\eta_1 \lambda_{1\rightarrow 0} & = &
\frac{1}{Z_M}[(K-1)\eta_0^{K-2}\eta_{00}\lambda_{00\rightarrow
1}],\\ \nonumber \eta_1 \lambda_{1\rightarrow 00} & = &
\frac{1}{Z_M}[(K-1)\eta_0^{K-2}\eta_{00}\lambda_{00\rightarrow
0}],\\ \nonumber \eta_0 \lambda_{0\rightarrow 1} & = &
\frac{e^{-y}}{Z_M}[(K-1)\eta_0^{K-2}\eta_{1}\lambda_{1\rightarrow
00}],\\ \nonumber \eta_0 \lambda_{0\rightarrow 00} & = &
\frac{1}{Z_M}[(K-1)\eta_0^{K-2}\eta_{1}\lambda_{1\rightarrow 0}],\\
\nonumber \eta_{00} \lambda_{00\rightarrow 1}& = &
\frac{e^{-y}}{Z_M}[(K-1)\eta_0^{K-1}\lambda_{0\rightarrow 00}],\\
\nonumber \eta_{00} \lambda_{00\rightarrow 0} & = &
\frac{1}{Z_M}[(K-1)\eta_0^{K-1}\lambda_{0\rightarrow 1}].
\end{eqnarray}

\begin{figure}
\includegraphics[width=10cm,angle=0]{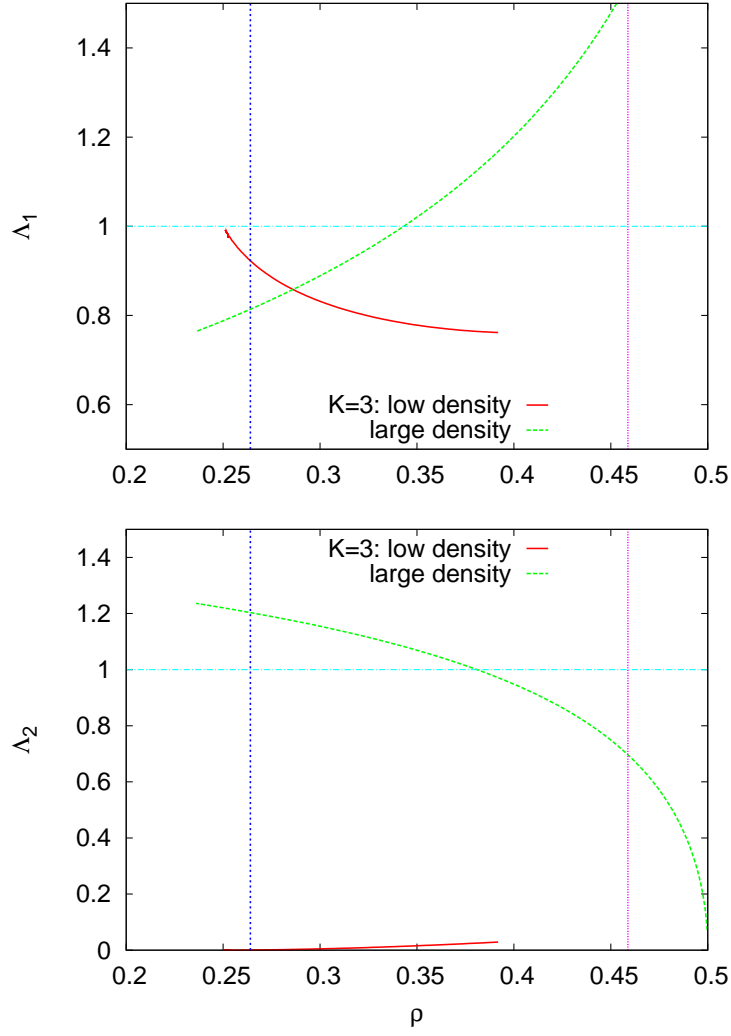}
\caption{(Color online) Checking stability of SP equations: typical behaviour of $\Lambda_1$ (upper picture) and $\Lambda_2$ (lower picture) in the low density (solid line) and large density (dashed line) region. Vertical dashed lines represent the minimum and maximum densities predicted by BP equations.}\label{stability-SP}
\end{figure}

If $\lambda_T$ is the dominant eigenvalue of $\mathbf{T}$ then the
second kind stability condition is $\Lambda_2= \lambda_T <1$. In
the present case, $\lambda_T=\frac{e^{-2y/3}}{Z_M}(K-1)\eta_0^{K-2}$
and the stability of SP solution is satisfied in the whole
low-density region, see Figure \ref{stability-SP}.

\subsubsection{Limit $\mu \to  - \infty$, $m \to 0$ ($y = m\mu$)}

In the high-density limit, equations simplify because $\eta_{00} = 0$
always and the matrix $\mathbf{M}$ reduces to the element
\begin{equation}
\partial \eta_1=-\frac{e^{-y}(1-\eta_1)^{K-2}}{Z_M}+\frac{e^{-y}(e^{-y}-1)(1-\eta_1)^{2K-3}}{Z_M^2}.
\end{equation}
with
\begin{equation}
Z_M=1+(e^{-y}-1)(1-\eta_1)^{K-1}.
\end{equation}
So for the first kind instability, the stable region satisfies
$\Lambda_1= (K-1)(\partial \eta_1)^2 <1$,

Similarly, equations for bug proliferation reduce to
\begin{eqnarray}
\eta_1 \lambda_{1\rightarrow 0}& =& \frac{1}{Z_M}
(1-\eta_1)^{K-1}[(K-1)\lambda_{0\rightarrow 1}],\\ \nonumber
(1-\eta_1) \lambda_{0\rightarrow 1} & =& \frac{e^{-y}}{Z_M}
(1-\eta_1)^{K-1}\left[(K-1)\eta_1(1-\eta_1)^{K-2}\lambda_{1\rightarrow
    0}\right].
\end{eqnarray}
so to have second kind stability we need $\Lambda_2=
\frac{(K-1)e^{-y/2}}{Z_M} (1-\eta_1)^{K-2} <1$. Numerical
solutions show that the SP solution is unstable in the whole
large-density region, see Figure \ref{stability-SP}.

\section{Population dynamics} \label{appPOP}

In the paper we had to use population dynamics two times; in the RS study of ER random graphs Eq. \ref{ERRSPr} and in 1RSB study
of random regular graphs Eq. \ref{1RSBPr}. Population dynamics is a way of solving these equations by representing a probability distribution
with a large population of variables. Let us describe how we solve the equation in the 1RSB case which is similar but more general than Eq. \ref{ERRSPr}.
This is the equation

\begin{eqnarray}
\mathcal{P}(\underline{r}^{i \rightarrow j})\propto \int  \prod_{k
\in \partial i \setminus j}d \mathcal{P}(\underline{r}^{k \rightarrow i})
e^{-y \Delta f_{k\rightarrow i}} \delta(\underline{r}^{i
\rightarrow j}-\mathcal{BP}).
\end{eqnarray}

We define a population of size $N_p$ with elements $\underline{r}^a, a=1,\ldots,N_p$. Each $\underline{r}^a$ is a probability vector
and frequency of a vector $\underline{r}$ in the population gives an estimate of $\mathcal{P}(\underline{r})$.

We start by a random initial population and in each step we update the population in the following way:

\begin{itemize}
\item select randomly members $a_1,\ldots,a_{K-1}$ from the population, here $K$ is node degree,
\item use $\underline{r}^{a_1},\ldots, \underline{r}^{a_{K-1}}$ to find $\underline{r}^{new}$ and $\Delta f_{cavity}$ according to BP equations,
\item with probability $\propto e^{-y \Delta f_{cavity}}$ replace a random member of population with $\underline{r}^{new}$,
\end{itemize}

After sufficiently large number iterations the population reaches a stationary state that can be used to obtain the interesting quantities.
For instance the generalized free energy $\Phi$ is given by

\begin{eqnarray}
y \Phi=  y \overline{ \Delta \Phi_i }-\frac{K}{2} y \overline{
\Delta \Phi_{ij} },
\end{eqnarray}

where the averages are taken over the population

\begin{eqnarray}
-y \overline{ \Delta \Phi_i } =  \ln \langle e^{-y \Delta f_i}
\rangle_{pop},\\ \nonumber -y \overline{ \Delta \Phi_{ij} }= \ln
\langle e^{-y \Delta f_{ij}} \rangle_{pop}.
\end{eqnarray}

Derivatives of $\Phi$ define the other average values.

\section{Two sub-problems of the mIS problem}\label{appPacCov}
In this Appendix we analyze the effects of the two constraints acting on the nodes of the graph separately.

\subsection{Packing constraint}
In this case we have only packing constraints $I_{ij}(\sigma_i,\sigma_j)$ which are satisfied if $\sigma_i  = 0 \vee \sigma_j = 0$. Then BP equations are
\begin{equation}
\nu_{i\rightarrow j}(\sigma_i)\propto \sum_{\sigma_{k\in \partial i
 \setminus j}} \prod_{k\in \partial i \setminus j}
I_{ik}(\sigma_i,\sigma_k)\nu_{k\rightarrow i}(\sigma_k)
\end{equation}
In random regular graphs (degree $K$) we take $\nu_{i\rightarrow
  j}(1)=r$ and the BP equations simplify into
\begin{equation}
r=\frac{e^{-\mu}(1-r)^{K-1}}{1+e^{-\mu}(1-r)^{K-1}}
\end{equation}

Then the free energy reads $\mu f=\mu \Delta f_i-\frac{K}{2}\mu \Delta f_{ij}=\mu \rho-s$,
\begin{eqnarray}
e^{-\mu \Delta f_i} & = & 1+e^{-\mu}(1-r)^K,\\ \nonumber
e^{-\mu \Delta f_{ij}} & = & 1-r^2,
\end{eqnarray}
and $\rho = \frac{e^{-\mu}(1-r)^K}{1+e^{-\mu}(1-r)^K}$.

\begin{figure}
\includegraphics[width=10cm,angle=0]{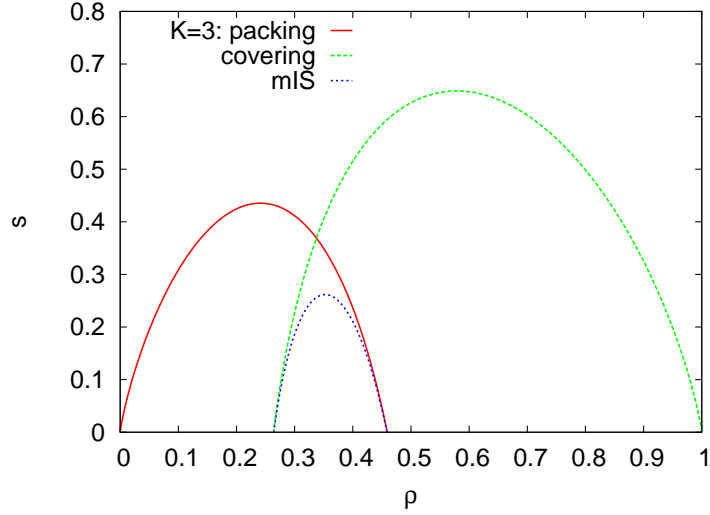}
\caption{(Color online) Comparing entropy of the packing (left curve), the
  covering (right curve) and mIS problems (middle curve) on random regular graphs of degree $K=3$.}\label{entropy-PC-K3}
\end{figure}

\subsection{Hyper-covering constraint}
The covering constraints $I_i(\sigma_i,\sigma_{\partial i})$ are satisfied if $\sigma_i + \sum_{j \in \partial i} \sigma_j >0$. The BP equations read
\begin{equation}
\nu_{i\rightarrow j}(\sigma_i,\sigma_{i\rightarrow j})\propto \sum_{\sigma_{k\rightarrow i}} \prod_{k\in \partial i\setminus j} I_{k}(\sigma_k,\sigma_{\partial k})\nu_{k\rightarrow i}(\sigma_k,\sigma_{k\rightarrow i})e^{-\mu \sigma_i}.
\end{equation}
In fixed degree graph we can define the cavity fields as for the full
problem $\{r_1^{i \to j}, r_0^{i \to j}, r_{00}^{i \to j} \}$ and we
get
\begin{eqnarray}
r_1 & =& \frac{e^{-\mu}}{e^{-\mu}+(1-r_{00})^{K-1}},\\ \nonumber
r_0 & =& \frac{(1-r_{00})^{K-1}-r_0^{K-1}}{e^{-\mu}+(1-r_{00})^{K-1}},\\ \nonumber
r_{00} & = & \frac{r_0^{K-1}}{e^{-\mu}+(1-r_{00})^{K-1}},
\end{eqnarray}

The free energy contributions are
\begin{eqnarray}
e^{-\mu\Delta f_i}=e^{-\mu}+(1-r_{00})^K-r_0^K,\\ \nonumber
e^{-\mu\Delta f_{ij}}=r_1^2+r_0^2+2r_1(r_0+r_{00}),
\end{eqnarray}
and $\rho=\frac{e^{-\mu}}{e^{-\mu}+(1-r_{00})^K-r_0^K}$.

The resulting curves for the BP entropy are displayed in the
representative case of $K=3$ in Fig. \ref{entropy-PC-K3}. The curve
for the packing problem is close to that of maximal-independent sets
for large density, whereas the hyper-covering (or conjugate lattice glass) curve
does the same for low densities. Together, the two curves define an
envelope that gives an upper bound for the BP entropy of the mIS's problem.

\section{Higher-order independent sets}\label{appmISn}
Higher order maximal-independent sets are important for applications
in computer science and economics. In particular, maximal-independent
sets of order 2 are the specialized Nash equilibria of the continuous
model of public goods game proposed by Bramoull\'e and Kranton in
Ref. \cite{BK07}.

In our CSP, a mIS of order $n$ is obtained just by imposing the
condition that each empty node has at least $n$ occupied neighbors.
For $n=1$ we recover the original mIS definition.
In general we write BP equations for the cavity fields
 $r_1^{i \to j} = R_{1,0}^{i \to j}$, $r_{>n-2}^{i \to j} =
\sum_{m > n-2} R_{0,m}^{i\to j}$, and $r_{>n-1}^{i \to j} = \sum_{m >
  n-1} R_{0,m}^{i \to j}$. On random regular graphs, they read
\begin{eqnarray}
r_1\propto e^{-\mu}r_{>n-2}^{K-1},\\ \nonumber
r_{>n-2}\propto \sum_{l=n-1}^{K-1} \binom{K-1}{l}r_1^l r_{>n-1}^{K-1-l},\\ \nonumber
r_{>n-1}\propto \sum_{l=n}^{K-1} \binom{K-1}{l}r_1^l r_{>n-1}^{K-1-l}.
\end{eqnarray}

Then the free energy reads

\begin{eqnarray}
\mu f=\mu\Delta f_i-\frac{K}{2} \mu\Delta f_{ij}=\mu \rho-s,\\ \nonumber
e^{-\mu\Delta f_i}=e^{-\mu}r_{>n-2}^K+\sum_{l=n}^{K} \binom{K}{l}r_1^l r_{>n-1}^{K-l},\\ \nonumber
e^{-\mu\Delta f_{ij}}=r_{>n-1}^2+2r_1 r_{>n-2},
\end{eqnarray}

and

\begin{eqnarray}
\rho= e^{-\mu}r_{>n-2}^K e^{\mu\Delta f_i}.
\end{eqnarray}

\begin{figure}
\includegraphics[width=10cm,angle=0]{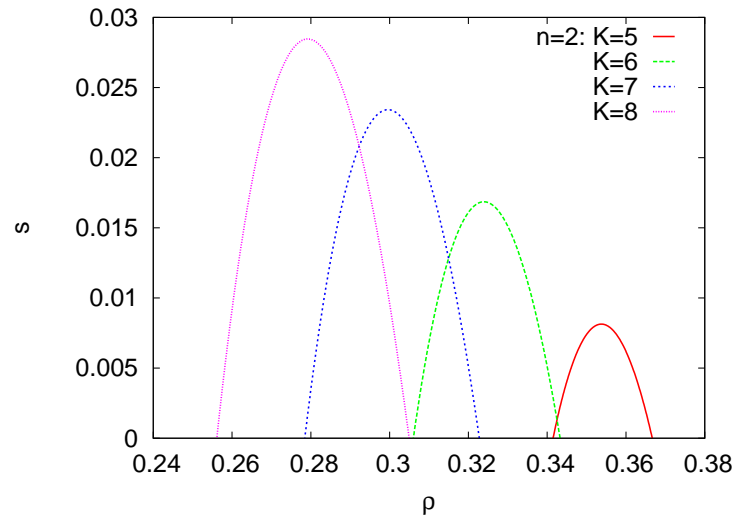}
\caption{(Color online) Entropy of order $n=2$ mIS's in some random regular graphs of degree $K=5-8$ (curves from right to left).}\label{fig_nmIS}
\end{figure}

In random regular graphs a maximal
independent set of order $n=2$ appears for the first time at $K=5$.
For larger degree values, the BP entropy is shown in Fig. \ref{fig_nmIS}.
It is worthy noting that the maximum entropy value increases for
larger degrees, in contrast to what happens in the $n=1$ case.

\end{document}